\documentclass[aps,eqsecnum,preprint,prd,showpacs]{revtex4}
\usepackage{graphicx}
\def\bold#1{\setbox0=\hbox{$#1$}%
     \kern-.025em\copy0\kern-\wd0
     \kern.05em\%\baselineskip=18ptemptcopy0\kern-\wd0
     \kern-.025em\raise.0433em\box0 }
\def\slash#1{\setbox0=\hbox{$#1$}#1\hskip-\wd0\dimen0=5pt\advance
         to\wd0{\hss\sl/\/\hss}}

\newcommand{\be}{\begin{equation}}
\newcommand{\ee}{\end{equation}}
\newcommand{\bea}{\begin{eqnarray}}
\newcommand{\eea}{\end{eqnarray}}
\newcommand{\nn}{\nonumber}
\newcommand{\dd}{\displaystyle}

\begin{document}
\begin{titlepage}
\addtolength{\jot}{10pt}

 \preprint{\vbox{\hbox{BARI-TH/06-542 \hfill}
                \hbox{October  2006\hfill} }}

\vspace*{1cm}

\title{\bf Spin effects
in  rare $B \to X_s \tau^+ \tau^-$ and   $B \to K^{(*)} \tau^+ \tau^-$  decays\\
in a single Universal Extra Dimension scenario}

\author{P. Colangelo$^a$, F. De Fazio$^a$, R. Ferrandes$^{a,b}$,  T.N. Pham$^c$ \\}

\affiliation{ $^a$ Istituto Nazionale di Fisica Nucleare, Sezione di Bari, Italy\\
$^b$ Dipartimento di Fisica,  Universit\'a  di Bari, Italy\\
$^c$ Centre de Physique Th\'eorique, \'Ecole Polytechnique, CNRS,\\
 91128 Palaiseau, France \\}

\begin{abstract}
We study the branching fractions  and the lepton polarization asymmetries of  rare
$B \to X_s \tau^+ \tau^-$ and
$B \to K^{(*)} \tau^+ \tau^-$  modes within the Standard Model and in the Appelquist-Cheng-Dobrescu
model, which is a New Physics  scenario with a single universal extra dimension.
In particular, we investigate the sensitivity of the observables to the radius $R$
of the compactified extra-dimension, which is the only new parameter in this New Physics model.
For the exclusive transitions we study the uncertainties due to the hadronic matrix elements
using two set of form factors, finding
that in many cases such uncertainties are small. In particular, we
show that the $\tau^-$ polarization asymmetries become free of hadronic uncertainties if
Large Energy relations for the form factors, valid for low momentum
transfer to the  lepton pair, are used. We also consider $K^*$ helicity fractions and their sensitivity to the
compactification parameter.
 \end{abstract}

\vspace*{1cm}

\maketitle
\end{titlepage}

\newpage
\section{Introduction}\label{sec:intro}
Among the various extensions of the Standard Model (SM) aimed at facing difficult problems such as
stability of the scalar sector under radiative corrections and  hierarchy,
 models with extra-dimensions (ED)
are receiving considerable attention since they can  hopefully
provide a unified framework for gravity and the other interactions
together with  a connection  with the string theory  \cite{rev}. A
special role is played by models where all the SM fields are
allowed to propagate in all available  dimensions. In these models
with so-called {\it universal} extra dimensions  (UED)   the extra
dimensions are compactified, and  above some compactification
scale $1/R$  the models are higher dimensional theories whose
equivalent description in four dimensions includes the ordinary
SM fields together with towers of their Kaluza-Klein (KK)
excitations and other fields having no  
Standard Model partners. In the case of a single universal extra dimension  a
model has been developed by Appelquist, Cheng and Dobrescu (ACD)
\cite{Appelquist:2000nn}, which presents the remarkable feature of
having only one new free parameter with respect to SM,  the
 radius $R$ of the compactified extra  dimension. The
masses of all the particles predicted in this model, together with
their interactions, are described in terms of the SM parameters
and of $\dd 1/R$, a noticeable economy in the theoretical
description, at least as far as the number of  fundamental
parameters is concerned. For example, as derived in
\cite{Appelquist:2000nn} and summarized in
\cite{Buras:2002ej,Buras:2003mk} and \cite{noi},  for  bosonic fields the masses of
the towers of KK excitations  are related to  the compactification
parameter according to the relation:  \be m_n^2=m_0^2+{n^2 \over R^2}
\,\,\,\,\,\,\,\,\,\, n=1,2,\dots \ee so that, for small values of $R$,
these particles, being more and more massive, decouple from the
low energy regime.

Another  feature of the ACD model is that in the equivalent
4-dimensional theory there is  conservation of the KK parity,
defined as $(-1)^j$, with $j$  the KK number. This implies the
absence of tree level contributions of Kaluza Klein states to
processes taking place at low energy, $\mu \ll 1/R$. For this
reason, flavour changing neutral current (FCNC) processes, which in SM
take contributions at loop level, are of particular interest,
since they are sensitive to loop contributions involving KK states
and therefore can be used to constrain their masses and couplings,
i.e. the compactification radius  \cite{Agashe:2001xt}. This
observation lead Buras and collaborators
 to compute in the ACD model  the
effective Hamiltonian of several FCNC processes, in particular in
the $b$ sector, namely $B_{d,s}$ mixing and $b \to s$ transitions
such as $b \to s \gamma$ and $b\to s \ell^+ \ell^-$ \cite{Buras:2002ej,Buras:2003mk}. For these
processes new  experimental data are accumulating, thanks to the
efforts at the B factories and at the Tevatron, so that they can
be used to constrain the extra dimension scenario. Moreover, in an
analysis of the exclusive $B \to K^{(*)} \ell^+ \ell^-$,  $B \to
K^{(*)} \nu \bar \nu$ and $B \to K^* \gamma$ modes it was also
shown that the uncertainty connected with  the hadronic matrix
elements does not mask the sensitivity to the compactification
parameter, and that current data, in particular the decay rates of
$B \to K^* \gamma$ and $B \to K^* \ell^+ \ell^-$ $(\ell=e,\mu)$
can provide a bound to the compactification parameter: $1/R \geq
300-400$ GeV \cite{noi}, which is similar to the bound obtained by
direct production of KK states at the Tevatron   \cite{Appelquist:2002wb}.
The modes $\Lambda_b \to \Lambda  \ell^+
\ell^-$ have  also been considered, in view of the possibility of
observing such processes at the hadron colliders \cite{aliev}.

 In this paper we  consider another set of observables in FCNC transitions,
namely those of the inclusive
 $B \to X_s \,\, + \,\,{ leptons}$  and exclusive  $B \to K^{(*)} \,\, + \,\, {leptons}$
 decay modes,  where the leptons are  $\tau^+  \tau^-$ pairs. At present, no
 experimental evidence of these modes is  available, yet. However, as first noticed  in
 \cite{Hewett:1995dk},  these processes are of great interest due  to the possibility of measuring
 lepton polarization asymmetries which  are sensitive to the structure of the interactions, so that they
 can be used to test  the Standard Model and its extensions. We  analyze in the single universal extra dimension scenario the $\tau$ polarization asymmetries in the inclusive $B \to X_s \tau^+ \tau^-$ transitions;
 we also consider  the exclusive
$B \to K^{(*)} \tau^+ \tau^-$ decays to investigate the impact  of the hadronic uncertainties.
Our result is that, although the sensitivity to extra dimension is milder than in other observables,
$\tau$ lepton polarization asymmetries, together with the decay rates,
can be used to provide additional constraints to the compactification parameter,   as well as to  confirm the Standard Model of the electroweak interactions.

Besides the $\tau$ polarization asymmetries, we  investigate another  observable,
 the fraction of longitudinal $K^*$ polarization in  $B \to K^{*} \ell^+ \ell^-$, for which
 a new measurement in two bins of momentum transfer to the lepton pair is avaliable  in case
 of $\ell=\mu, e$.   The dependence of  this quantity on the compactification parameter, for
  $B \to K^{*} \tau^+ \tau^-$ and in the case of light leptons, together with the fraction of transverse $K^*$ polarization in the same modes,    provides us with
  another possibility to constrain  the universal extra dimension scenario.

The plan of the paper is the following.   In Section \ref{sec:hamiltonian} we recall the effective Hamiltonian
inducing $b \to s \ell^+ \ell^-$ transitions in SM and in the ACD model, together with the
definition of the polarization asymmetries we consider in our study.
In Sections  \ref{sec:inclusive} and   \ref{sec:BK},\ref{sec:BKstar} we analyze $\tau^-$ polarization asymmetries in inclusive $B \to X_s \tau^+ \tau^-$ and exclusive $B\to K^{(*)} \tau^+ \tau^-$ decays, paying  particular
attention to the hadronic uncertainties in the exclusive modes.
  In Section \ref{sec:BKstarhelicity} we study the helicity fractions of $K^*$ in $B\to K^{*} \ell^+ \ell^-$
($\ell=e, \mu)$, discussing the experimental data, and  in  $B\to K^{*} \tau^+ \tau^-$.
Finally, in Section \ref{sec:conclusions} we  draw our conclusions.

\section{ Effective Hamiltonian} \label{sec:hamiltonian}

In the Standard Model  the transitions $b \to s \ell^+ \ell^-$ are described by the effective
$\Delta B=-1$, $\Delta S=1$ Hamiltonian
 \begin{equation} H_W\,=\,4\,{G_F
\over \sqrt{2}} V_{tb} V_{ts}^* \sum_{i=1}^{10} c_i(\mu) O_i(\mu)
\label{hamil}
\end{equation}
\noindent
obtained by a  renormalization group evolution,
including QCD corrections,  from the electroweak scale down to $\mu\simeq m_b$.
$G_F$ is the Fermi constant and
$V_{ij}$ are elements of the Cabibbo-Kobayashi-Maskawa matrix \cite{buchalla}.
The local operators
$O_i$, written in
terms of quark and gluon fields,  read as follows:
\begin{eqnarray}
O_1&=&({\bar s}_{L \alpha} \gamma^\mu b_{L \alpha})
      ({\bar c}_{L \beta} \gamma_\mu c_{L \beta}) \nonumber \\
O_2&=&({\bar s}_{L \alpha} \gamma^\mu b_{L \beta})
      ({\bar c}_{L \beta} \gamma_\mu c_{L \alpha}) \nonumber \\
O_3&=&({\bar s}_{L \alpha} \gamma^\mu b_{L \alpha})
      [({\bar u}_{L \beta} \gamma_\mu u_{L \beta})+...+
      ({\bar b}_{L \beta} \gamma_\mu b_{L \beta})] \nonumber \\
O_4&=&({\bar s}_{L \alpha} \gamma^\mu b_{L \beta})
      [({\bar u}_{L \beta} \gamma_\mu u_{L \alpha})+...+
      ({\bar b}_{L \beta} \gamma_\mu b_{L \alpha})] \nonumber \\
O_5&=&({\bar s}_{L \alpha} \gamma^\mu b_{L \alpha})
      [({\bar u}_{R \beta} \gamma_\mu u_{R \beta})+...+
      ({\bar b}_{R \beta} \gamma_\mu b_{R \beta})] \nonumber \\
O_6&=&({\bar s}_{L \alpha} \gamma^\mu b_{L \beta})
      [({\bar u}_{R \beta} \gamma_\mu u_{R \alpha})+...+
      ({\bar b}_{R \beta} \gamma_\mu b_{R \alpha})] \nonumber \\
O_7&=&{e \over 16 \pi^2} \left[ m_b ({\bar s}_{L \alpha}
\sigma^{\mu \nu}
     b_{R \alpha}) +m_s ({\bar s}_{R \alpha}
\sigma^{\mu \nu}
     b_{L \alpha}) \right] F_{\mu \nu} \nonumber \\
O_8&=&{g_s \over 16 \pi^2} m_b \Big[{\bar s}_{L \alpha}
\sigma^{\mu \nu}
      \Big({\lambda^a \over 2}\Big)_{\alpha \beta} b_{R \beta}\Big] \;
      G^a_{\mu \nu} \nonumber\\
O_9&=&{e^2 \over 16 \pi^2}  ({\bar s}_{L \alpha} \gamma^\mu
     b_{L \alpha}) \; {\bar \ell} \gamma_\mu \ell \nonumber \\
O_{10}&=&{e^2 \over 16 \pi^2}  ({\bar s}_{L \alpha} \gamma^\mu
     b_{L \alpha}) \; {\bar \ell} \gamma_\mu \gamma_5 \ell
\label{eff}
\end{eqnarray}
\noindent ($\alpha$, $\beta$ are colour indices, $\displaystyle
b_{R,L}={1 \pm \gamma_5 \over 2}b$, and $\displaystyle \sigma^{\mu
\nu}={i \over 2}[\gamma^\mu,\gamma^\nu]$; $e$ and $g_s$ are the
electromagnetic and the strong coupling constant, respectively,
and $F_{\mu \nu}$ and $G^a_{\mu \nu}$ in $O_7$ and $O_8$ denote
the electromagnetic and the gluonic field strength tensor). $O_1$
and $O_2$ are current-current operators, $O_3,...,O_6$   QCD
penguins, $O_7$ and $O_8$  magnetic penguins,  $O_9$ and $O_{10}$
 semileptonic  penguin operators.

In the ACD model, no  operators other than those in eq.(\ref{eff})
are found to
 contribute to $b \to s \ell^+ \ell^-$ transitions  \cite{Buras:2002ej,Buras:2003mk}.
 The model consists in a minimal extension of SM in $4+1$ dimensions,  with the extra dimension
 compactified to the orbifold $S^1/Z_2$. The fifth coordinate $y$ runs from $0$ to $2 \pi R$;  the points $y=0$, $y=\pi R$ are fixed points of the orbifold, and the boundary conditions of the fields at these points determine their Kaluza Klein mode expansion. Under parity transformation $P_5: y \to -y$ fields existing
  in the Standard Model are even, and their zero modes in the KK expansion are interpreted as the ordinary SM fields. On the other hand, fields  absent in SM are odd under $P_5$,
 so they  do not present zero modes.

The effect of the new states predicted in the extra-dimension
model only consists in a modification of  the Wilson coefficients
$c_{1-10}$ in (\ref{hamil});  in particular,  the coefficients acquire a dependence
on the new parameter of the model, the compactification
radius $R$.  For large values of $1/R$, due to decoupling of the
new states predicted in the ACD model,   the Wilson coefficients reproduce
those obtained in the Standard Model, so that the Standard Model
phenomenology is  recovered in that limit. Actually, the
coefficients can be generally expressed in terms of functions
$F(x_t,1/R )$, with $x_t=\displaystyle{ m_t^2 \over M_W^2}$ and $m_t$
the top quark mass. Such functions generalize their SM
analogues $F_0(x_t)$ according to: \be
F(x_t,1/R)=F_0(x_t)+\sum_{n=1}^\infty F_n(x_t,x_n) \,, \label{fxt}
\ee
 with
$x_n=\displaystyle{ m_n^2 \over M_W^2}$, $m_n=\displaystyle{n
\over R}$. As discussed in \cite{Buras:2002ej,Buras:2003mk}, the
sum in eq. (\ref{fxt}) is  finite in all cases as a
consequence of a generalized GIM mechanism,  and fulfills the
condition $F(x_t,1/R) \to F_0(x_t)$ when $R \to 0$.
However, as far as
$1/R$ is kept of the order of a few hundreds of GeV, the
coefficients differ from the Standard Model value: in particular,
$c_{10}$ is enhanced and $c_7$ is suppressed. This implies   that
deviations could be seen in various observables in modes induced
by the transition $b \to s \ell^+ \ell^-$, namely branching
fractions and  forward-backward lepton asymmetry
 \cite{Buras:2002ej,Buras:2003mk,noi}.

In the following we only consider the contribution of the operators $O_7$, $O_9$ and $O_{10}$.
We do not take into account the one-loop contribution of the four-quark operators $O_1-O_6$,
which is small \cite{Grinstein:1988me},
and the long distance contribution associated with the real $\bar c c$
resonances as intermediate states decaying to lepton pairs \cite{Deshpande:1988bd},
which can be removed by appropriate kinematical cuts. 
The Wilson coefficients $c_7$, $c_9$ and $c_{10}$ are real;
their expressions in ACD  can be found in Refs.\cite{Buras:2002ej,Buras:2003mk} and \cite{noi}. However, in the formulae
we  provide below for the polarization asymmetries we  consider the general case of
complex coefficients,  with the aim of  explicitely showing why some asymmetries are suppressed with respect to others.

To compute lepton polarization asymmetries  for $B$ decays  in  $\tau$ leptons  we consider the spin vector $s$ of  $\tau^-$, with  $s^2=-1$ and $k_1 \cdot s=0$,   $k_1$ being
the $\tau^-$ momentum.  In the rest frame of the $\tau^-$ lepton
three orthogonal
unit vectors: $e_L$, $e_N$ and $e_T$  can be defined, corresponding to the
longitudinal $s_L$, normal $s_N$ and transverse $s_T$ polarization vectors:

\bea s_L &=& (0,e_L)=\left( 0, {\vec k_1 \over |\vec k_1|}
\right) \nonumber \\
s_N &=& (0,e_N)=\left( 0, {\vec p^\prime \times \vec k_1 \over
|\vec p^\prime \times \vec k_1|}
\right)  \label{spinsrf} \\
s_T &=& (0,e_T)=(0,e_N \times e_L) \,\,\, ;  \nonumber  \eea
\noindent
in eq.(\ref{spinsrf}) $\vec p^\prime$ and $\vec k_1$ are respectively  the strange quark (or $K,
K^* $ meson in the exclusive decays) and the $\tau^-$  three-momenta in the rest
frame of the lepton pair.  Choosing  the
$z$-axis directed as the $\tau^-$ momentum in the rest frame of the
lepton pair: $k_1=(E_1,0,0,|\vec k_1|)$ and boosting the spin vectors $s$
in (\ref{spinsrf}) in the rest frame of the lepton pair,  the normal and transverse polarization vectors
$s_N,s_T$ remain unchanged: $s_N=(0,1,0,0)$ and
$s_T=(0,0,-1,0)$, while the longitudinal polarization
vector becomes:
\be s_L={1 \over m_\tau}(|\vec k_1|,0,0,E_1) \,\,\, . \label{sl} \ee
For each value of the squared momentum transfered to the lepton pair,  $q^2$,
the polarization asymmetry for the negatively charged  $\tau^-$ lepton  is
defined as: \be {\cal A}_A(q^2)=\displaystyle{ {d \Gamma \over dq^2}(s_A)-{d
\Gamma \over dq^2}(-s_A) \over {d \Gamma \over dq^2}(s_A)+{d
\Gamma \over dq^2}(-s_A) } \label{def-pol} \ee with $A=L,T$ and $N$; 
these are the observables we consider  in the case of  inclusive
$B \to X_s \tau^+ \tau^-$ and exclusive $B \to K^{(*)}  \tau^+ \tau^-$ decays.

\section{Lepton polarization asymmetries in $B \to X_s \tau^+ \tau^-$} \label{sec:inclusive}

The expressions for the longitudinal (L), transverse (T) and normal (N) $\tau^-$ polarization asymmetries
in $B \to X_s \tau^+ \tau^-$ can be derived from the transition  amplitude
\bea {\cal M}={G_F \over \sqrt{2}} V_{tb} V_{ts}^* {\alpha \over
\pi} \, &\Big[&c_9(\mu, 1/R) \, {\bar s}_L \gamma_\mu b_L {\bar
\ell} \gamma_\mu \ell + c_{10}(\mu, 1/R) \, {\bar s}_L \gamma_\mu
b_L {\bar \ell} \gamma_\mu \gamma_5 \ell \nn \\ &-& 2 c_7(\mu,
1/R)\, {q^\nu \over q^2} \, \left[ m_b
 {\bar s}_L i \sigma_{\mu \nu} b_R + m_s {\bar s}_R i \sigma_{\mu \nu} b_L \right]
 {\bar \ell}
\gamma_\mu \ell \Big]  \label{ampl} \eea
 stemming from the effective Hamiltonian (\ref{hamil}), so that
the measurements of such asymmetries can be used to constrain
 the Wilson coefficients   \cite{Hewett:1995dk,Kruger:1996cv}.
The three   asymmetries read as follows:
\be  {\cal A}_L(b \to s \tau^+ \tau^-) =2
 q^4 \sqrt{1-{4 m_\tau^2 \over q^2}} {\rm Re}\left\{c_{10}
 \left[6 c_7^* h(m_b^2,m_s^2,q^2)+c_9^* f(m_b^2,m_s^2,q^2) \right]
 \right\}{1 \over d(q^2)} \nn \\ \label{aLincl} \ee
\bea &&{\cal A}_T(b \to s \tau^+ \tau^-) =-{3 \over 2}\pi m_\tau
\sqrt{q^2} \lambda^{1/2} \nn  \\
&&\left\{(m_b^2-m_s^2)\left[4c_7^2 \, (m_b^2-m_s^2)-q^2\, {\rm
Re}[c_{10}(2 c_7^*+c_9^*)] \right]+c_9^2 \, q^4+4q^2(m_b^2+m_s^2)
{\rm Re}[ c_7^* c_9] \right\} {1 \over d(q^2)} \nonumber
 \\ \label{aTincl} \eea
\be  {\cal A}_N(b \to s \tau^+ \tau^-) ={3 \over 2}\pi m_\tau
\sqrt{1-{4 m_\tau^2 \over q^2}} q^2 \sqrt{q^2} \lambda^{1/2} {\rm
Im}[(2c_7^*(m_b^2+m_s^2)+q^2 \, c_9^*)c_{10}]{1 \over d(q^2)} \nn \\
 \label{aNincl}
\ee
for  longitudinally, transversely and normally polarized $\tau^-$ leptons, respectively,
with  the function $d$  given by
\bea
 d(s)&=&(2 m_\tau^2+s) \Big\{ s
 f(m_b^2,m_s^2,s)(c_9^2+c_{10}^2)+12s \,h(m_b^2,m_s^2,s){\rm
 Re}[c_7 c_9^*] \nonumber \\
&& -4 c_7^2
\left[(m_b^2+m_s^2)(s^2-(m_b^2-m_s^2)^2-h(m_b^2,m_s^2,s))+12s
m_b^2 m_s^2 \right] \Big\} \nonumber \\
&&-12s^2 c_{10}^2 m_\tau^2(m_b^2+m_s^2-s) \eea 
and
\bea
h(m_b^2,m_s^2,s)&=&(m_b^2-m_s^2)^2-s(m_b^2+m_s^2) \,\,\,\, \\
f(m_b^2,m_s^2,s)&=&(m_b^2-m_s^2)^2+(m_b^2+m_s^2)s-2s^2 \,\,\, ; \eea 
$\lambda(m_b^2,m_s^2,q^2)$ is the triangular function. 

Looking at eq.(\ref{aNincl}) one sees that the asymmetry for normally polarized $\tau^-$ leptons is 
small in SM, and remains small even in ACD. As a matter of fact, in SM only the Wilson coefficient
$c_9$ has an imaginary part if the (small) short-distance contribution of the four quark operators 
$O_1$ and $O_2$ is considered, so that $c_9$ must be substituted by a $c_9^{eff}$,
or if the  long-distance contribution of $\bar c c$ resonances
is included.  As we have mentioned above, we  neglect such terms in our analysis. Therefore,
${\cal A}_N$ is an interesting observable to investigate models producing  imaginary parts
to the coefficients: in our framework, it is expected to be nearly vanishing, a
 behaviour  also expected in the exclusive modes. 
 
\begin{figure}[t]
\begin{center}
\includegraphics[width=0.45\textwidth] {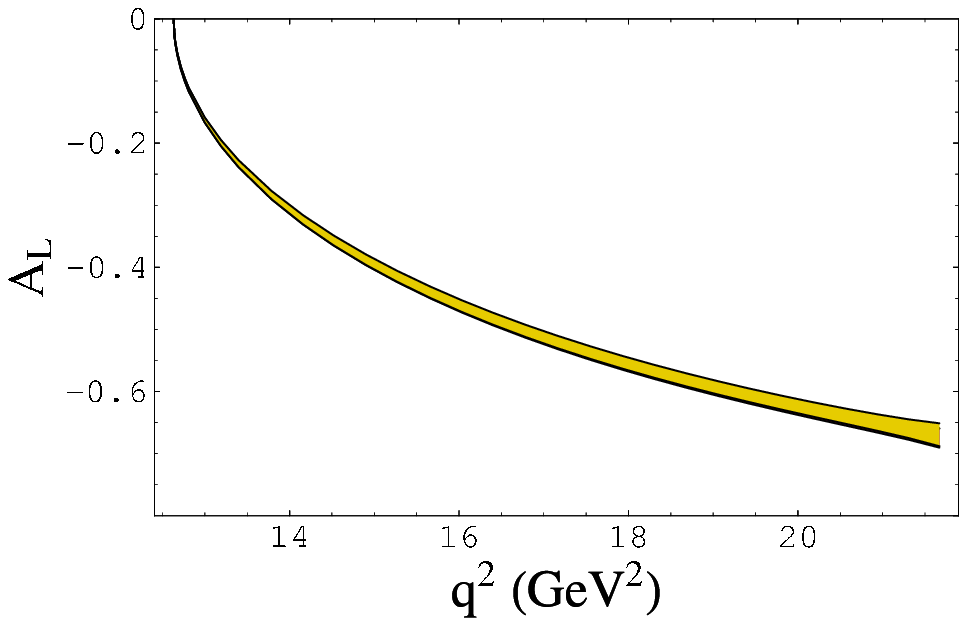} \hspace{0.5cm}
 \includegraphics[width=0.45\textwidth] {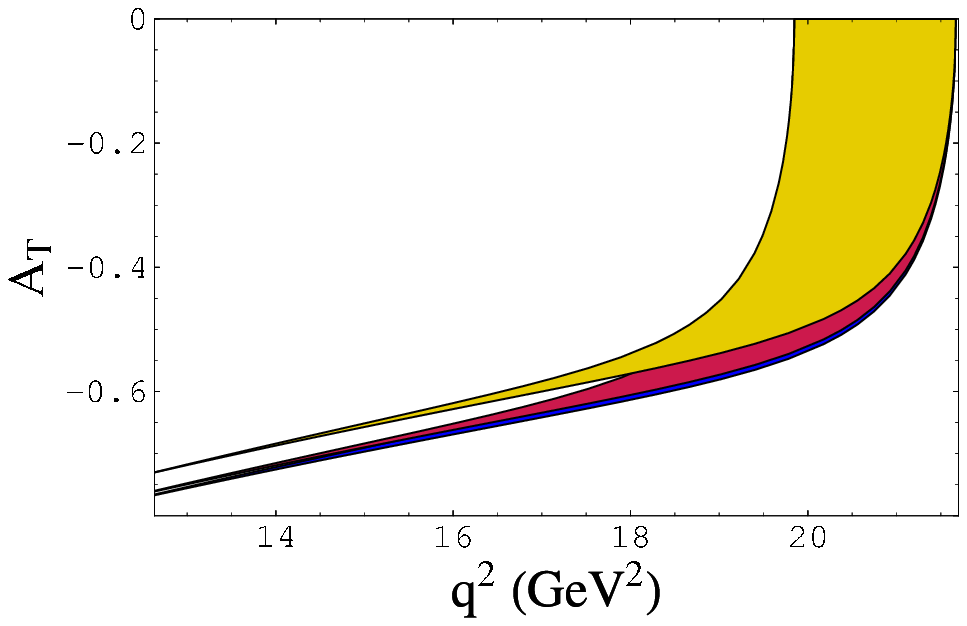}
\end{center}
\caption{\baselineskip=15pt Longitudinal (left) and transverse
(right) $\tau^-$ polarization asymmetry in the inclusive mode $B \to X_s \tau^+
\tau^-$.
 The  dark (blue) region is obtained in  SM;
 the intermediate (red) one for $1/R=500$ GeV, the light (yellow) one for $1/R=200$ GeV.} \vspace*{1.0cm}
\label{asymLTincl}
\end{figure}

Let us concentrate on  ${\cal A}_{L,T}$.
 In  fig.\ref{asymLTincl}  we depict the result obtained for these two observables
in the Standard Model,  using $m_s
=0.145\pm 0.015$ GeV  \cite{ms,PDG} and $m_b=4.8 \pm 0.2$ GeV \cite{PDG}, and in the ACD model for two values 
of the compactification parameter, $\dd {1 \over R}=500$ and $200$ GeV. 
%
In the longitudinal polarization asymmetry ${\cal A}_{L}$ the effect of the extra dimension is tiny for all values of the
momentum transfer $q^2$, at least for a compactification parameter in the range  we have chosen,
$\dd{1 \over R} \geq 200$ GeV. Sensitivity to $\dd{1 / R}$ is relatively higher for the transverse
 asymmetry, mainly in the low $q^2$ region where this observable reaches  the largest
(absolute) value. At  higher $q^2$ the uncertainty related to the input parameters,   the
strange and beauty quark masses, obscures the sensitivity to the compactification radius.

The situation is different  in the exclusive modes, where however one has to take into account
the hadronic uncertainties.  As  we shall see below, it is reassuring that
such uncertainties remain under control
in  particular kinematic configurations. 

\section{Exclusive $B \to K \tau^+ \tau^-$ mode}\label{sec:BK}

The description of the exclusive $B \to K^{(*)} \tau^+ \tau^-$ modes
involves  hadronic matrix elements of the effective Hamiltonian (\ref{hamil}).
For  $B \to K \tau^+ \tau^-$ three form factors are needed to parameterize  such
matrix elements:
\bea
<K(p^\prime)|{\bar s} \gamma_\mu b |B(p)>&=&(p+p^\prime)_\mu
F_1(q^2) +{M_B^2-M_K^2 \over q^2} q_\mu \left
(F_0(q^2)-F_1(q^2)\right ) \nn \\
<K(p^\prime)|{\bar s}\; i\;  \sigma_{\mu \nu} q^\nu b |B(p)>&=&
\Big[(p+p^\prime)_\mu q^2 -(M_B^2-M_K^2)q_\mu\Big] \; {F_T(q^2)
\over M_B+M_K} \hskip 3 pt , \label{ft}
\eea
where  $q=p-p^\prime$ and the condition $F_1(0)=F_0(0)$ is imposed.
As discussed  in \cite{noi}, to account for the uncertainties in the hadronic matrix elements we
use two sets of form factors, one (denoted as set A) obtained by three-point QCD sum rules
\cite{Colangelo:1995jv}, and another one (set B) obtained by light-cone sum rules  \cite{Ball:2004rg}.
The differences in the results obtained using the two sets  represents  an indication of  the  error
related to the hadronic uncertainty.

The differential $B \to K \tau^+ \tau^-$ decay rate can be written
as: \be {d\Gamma(q^2) \over d q^2}= {G_F^2|V_{tb}V_{ts}^*|^2
\alpha^2 \over 2^{9} \pi^5} {\lambda^{1/2}(M_B^2,M_{K}^2,q^2)
\over M_B^3}\sqrt{1-{4 m_\tau^2 \over q^2}}{1 \over 3  q^2}p(q^2)
\label{dGBK}\ee where: \be p(s)=6m^2_\tau(M_B^2-M_K^2)^2|b(s)|^2+
\lambda(M_B^2,M_{K}^2,q^2)\left[ (2 m_\tau^2+s)|c(s)|^2-(4
m_\tau^2-s)|a(s)|^2 \right] \label{denT} \ee and the following
combinations of form factors and Wilson coefficients have been
introduced:
\bea a(s)&=& c_{10} F_1(s) \nonumber \\
b(s) &=&  c_{10}F_0(s)\label{abc} \\
c(s) &=& c_9 F_1(s)- 2 (m_b+m_s) c_7 { F_T(s) \over M_B+M_K}
\,\,\,. \nonumber \eea

Integrating (\ref{dGBK}) over all  the range of  momentum transfer:
$4 m_\tau^2 \leq q^2 \leq (M_B-M_K)^2$, and
 using  the values reported by the PDG
\cite{PDG} for CKM matrix elements and  $B^0$ lifetime,
we obtain in the Standard Model
the branching fraction
\be
BR(B \to K \tau^+ \tau^-)= \Bigg \{
\begin{array}{c}(0.6 \pm 0.1) \,\,\, 10^{-7}  \,\,\, {\rm (set \,\, A)} \\ (1.6 \pm 0.3)\,\,\, 10^{-7}  \,\,\, {\rm (set \,\, B) \,\,\, .} \end{array}
\ee
The difference in results obtained using  the two sets of form factors was already noticed
in \cite{noi} for $B \to K \mu^+ \mu^-$; in that case,  the present experimental results
for the branching ratio
are compatible with both the predictions based on  set A and B.

\begin{figure}[t]
\begin{center}
\includegraphics[width=0.45\textwidth] {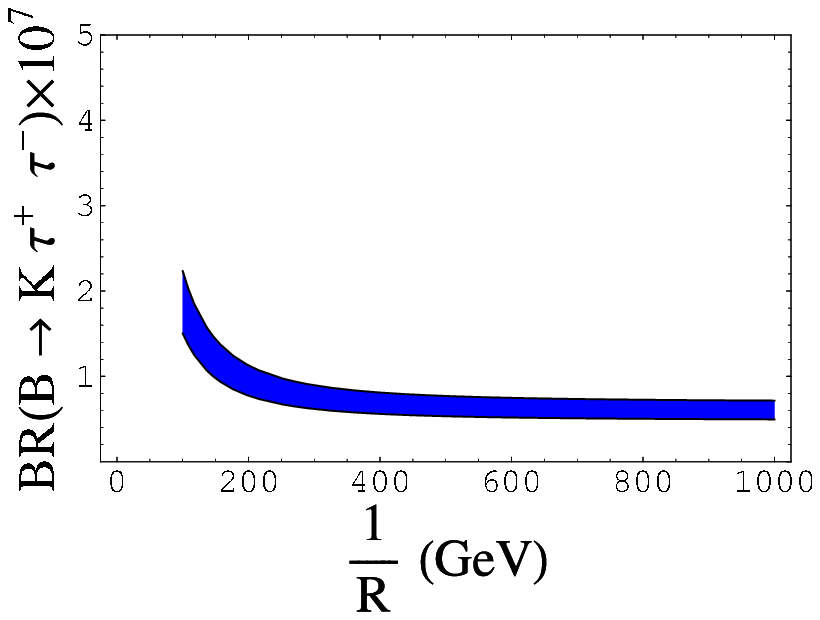} \hspace{0.5cm}
 \includegraphics[width=0.45\textwidth] {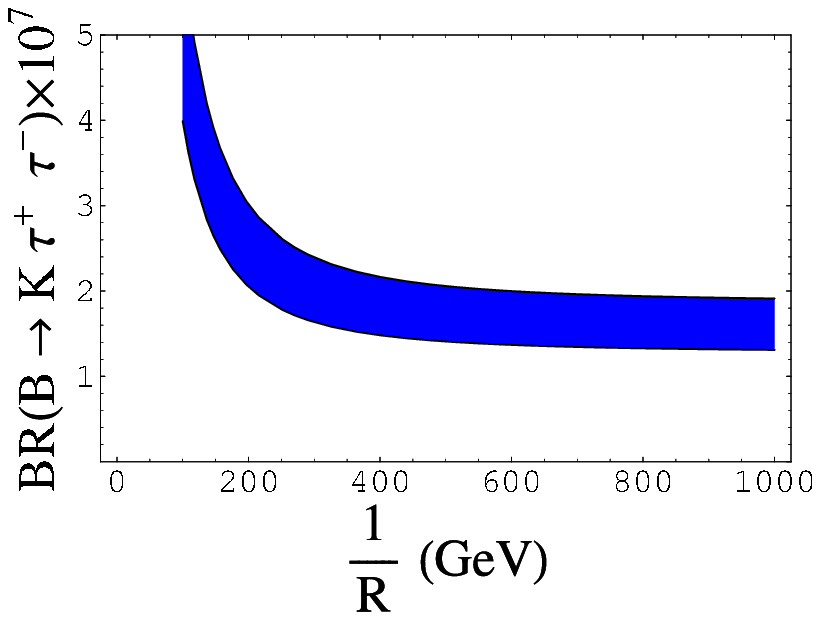}\\
\end{center}
\caption{\baselineskip=15pt  $BR(B \to K \tau^+ \tau^-)$ versus $\dd 1/R$
 obtained using  set A   (left) and B  (right) of form factors.} \vspace*{1.0cm}
\label{brktau}
\end{figure}

In fig.\ref{brktau} we depict the branching fraction of $B
\to K \tau^+ \tau^-$ versus the compactification parameter $\dd 1/R$ for the two set of form factors.
The modification induced by extra-dimensions would be observable for small values of $\dd 1/R$: in particular, a branching fraction exceeding $2 \times 10^{-7}$ would provide us with the
bound $\displaystyle {1 \over R} \leq 300$ GeV using set B.

Coming to the $\tau^-$ polarization asymmetries, considered in the Standard Model and
in some extensions in  \cite{Geng:1996az}, the expressions
of the longitudinal ${\cal A}_L$ and transverse ${\cal A}_T$
asymmetry can be given in terms of the combinations of Wilson
coefficients and form factors in eq.(\ref{abc}) and of the
function in (\ref{denT}): \be {\cal A}_L(q^2) = {2 {\rm Re}[a(q^2)
c^*(q^2)] \over p(q^2)} \sqrt{1-{4 m_\tau^2 \over q^2}} q^2 \,
\lambda(M_B^2,M_{K}^2,q^2)   \label{aLK} \ee
\noindent and
 \be {\cal
A}_T(q^2) = {3 \over 2} \pi m_\tau (M_B^2-M_K^2) \sqrt{q^2} \,
\lambda^{1/2}(M_B^2,M_{K}^2,q^2) { {\rm Re}[b(q^2) c^*(q^2)] \over
p(q^2)}  \label{aTK} \,\,\,\, . \ee

\begin{figure}[t]
\begin{center}
\includegraphics[width=0.45\textwidth] {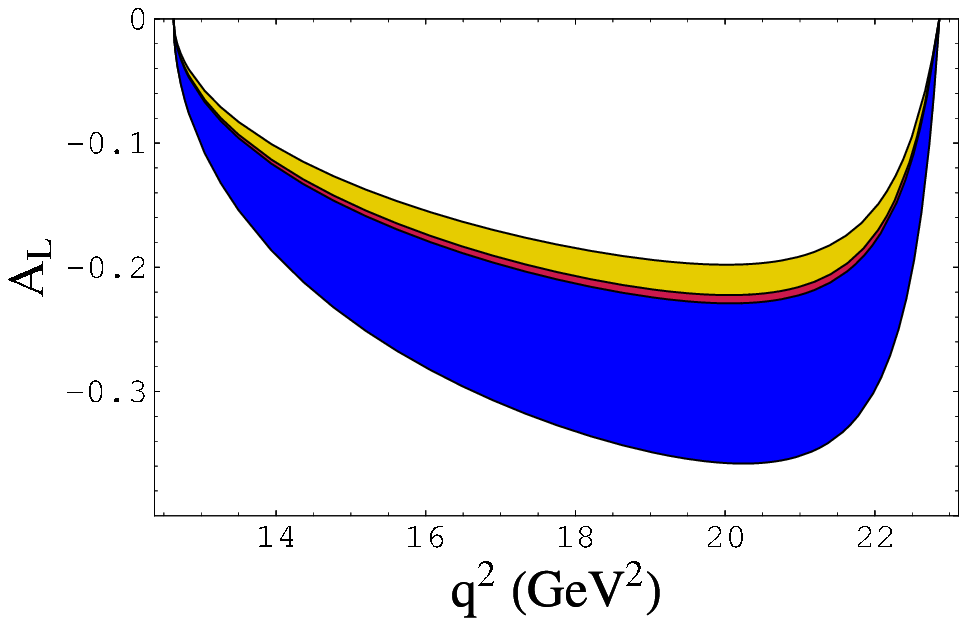} \hspace{0.5cm}
 \includegraphics[width=0.45\textwidth] {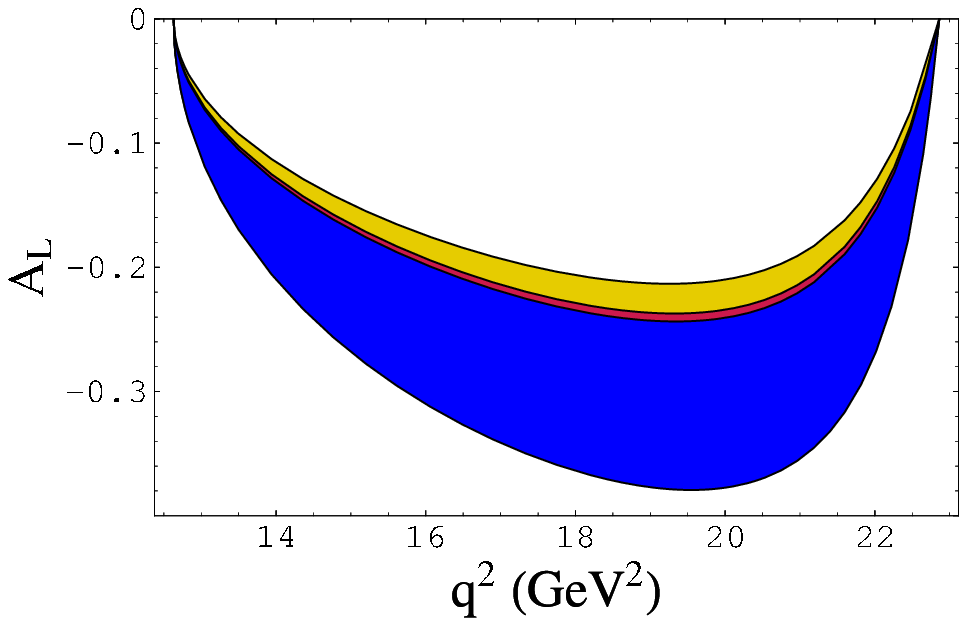}\\ \vspace{0.5cm}
 \includegraphics[width=0.45\textwidth] {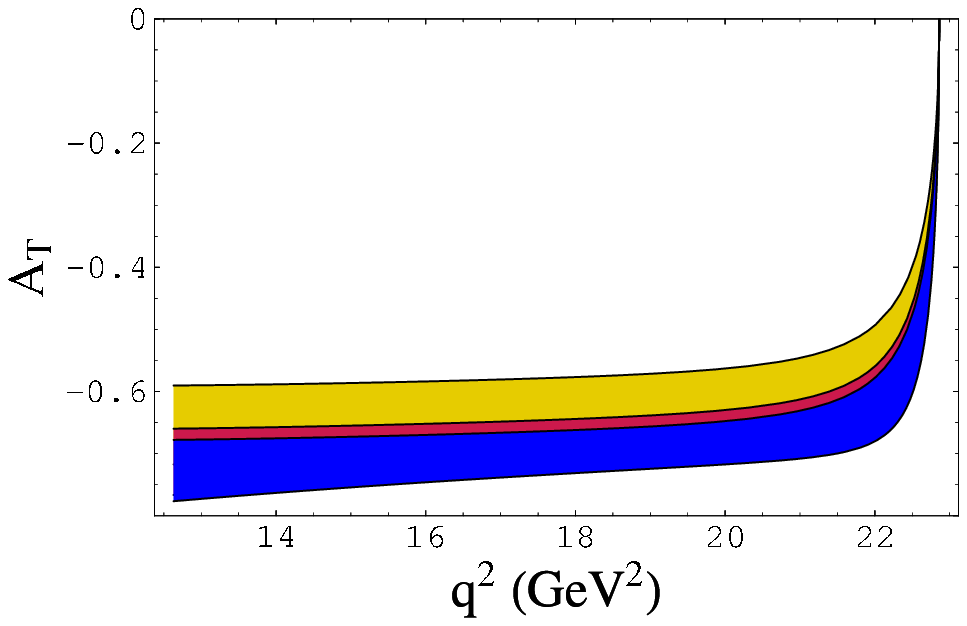} \hspace{0.5cm}
 \includegraphics[width=0.45\textwidth] {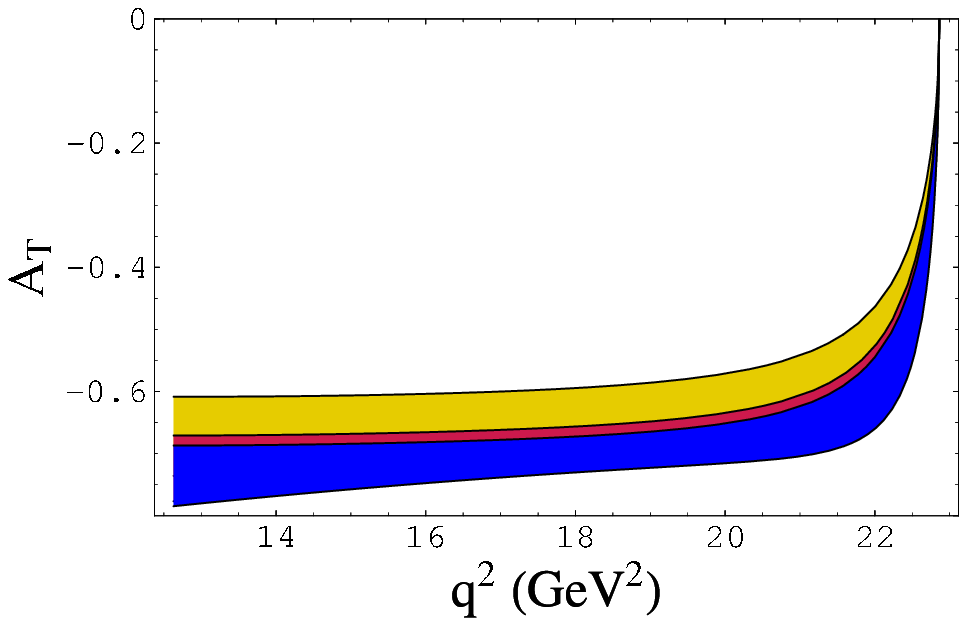}
\end{center}
\caption{\baselineskip=15pt Longitudinal (up) and transverse (down) $\tau^-$ polarization
asymmetry in $B \to K \tau^+ \tau^-$
 obtained using  set A   (left) and B  (right) of form factors.
 The  dark (blue) region is obtained in  SM,
 the intermediate (red) one for $1/R=500$ GeV, the light (yellow) one for $1/R=200$ GeV.} \vspace*{1.0cm}
\label{asymLk}
\end{figure}

 In fig.\ref{asymLk} we show  the results
obtained  in the Standard Model and for two values of the
compactification parameter: $\displaystyle {1\over R}=500$ and
$200$ GeV.
%
%
It is interesting to observe that the hadronic uncertainty for the polarization asymmetries
is very similar for the two sets of form factors. The longitudinal polarization asymmetry
${\cal A}_L$, which vanishes
at the extremes of the $q^2$ range, presents the maximal deviation from zero at $q^2 \simeq 20$ GeV$^2$, where $-0.38 \leq {\cal A}_L \leq -0.22$ in the Standard Model. The effect of extra dimension
is  modest, since ${\cal A}_L$ is shifted at most at  ${\cal A}_L\simeq-0.18$
for the smallest  value  $\dd {1 \over R}=200$ GeV of the compactification parameter.

The transverse polarization asymmetry ${\cal A}_T$, which has the largest (absolute) value in the small
$q^2$ range, is more sensitive to ED: for $\displaystyle {1 \over R}=200$ GeV it systematically deviates from
the Standard Model value
practically in the whole  $q^2$ kinematical range,  and this is
an experimentally accessible effect.  The conclusion is that  small values of the
compactification parameter  simultaneously induce in $B \to K \tau^+ \tau^-$ mode an increase of
the branching fraction and a  decrease of the transverse polarization asymmetry
in a broad range of momentum transfer. Observation of such a correlation  between the two observables
is a challenge for experimental investigations.

Let us elaborate more on $B \to K \tau^+ \tau^-$
polarization asymmetries.   In deriving their expressions it is possible
to exploit relations among form factors that can be obtained in the large energy
limit of the final meson  for $B$ meson decays to a light hadron \cite{scet}. For
 $B \to P$ transitions, with $P$ being a light pseudoscalar
meson (in our case a kaon) and for large values of the light meson energy
$E$ in the $B$ meson rest frame:
$E={M_B\over 2}(1+{M_K^2 \over M_B^2}-{q^2\over M_B^2})$,
the three form factors $F_1$, $F_0$ and $F_T$
 can be related to a single hadronic function $\xi_P$:
 \bea
F_1(q^2)&=& \xi_P(E) \nonumber \\
F_0(q^2) &=& {2 E \over M_B} \xi_P(E) \\
F_T(q^2)&=& -{M_B+M_K \over M_B} \xi_P(E) \,\,\ .\nonumber \label{scetP} \eea
The consequence is that, in such a limit,
the longitudinal and transverse polarization asymmetries become  independent of  form factors, since
the hadronic function $\xi_P$ cancels out in the ratio defining  ${\cal A}_L$ and  ${\cal A}_T$:
\bea {\cal A}_L^{LE}(q^2)&=&\sqrt{1-{4 m^2_\tau \over q^2}} q^2  {2 {\rm Re}[ c^*_{10} M_B (c_9
M_B +2 c_7 m_b)] \over \left[|c_{10}|^2 M_B^2 +|c_9 M_B +2 c_7 m_b|^2
\right](2 m_\tau^2+q^2)} \label{alkscet} \\
{\cal A}_T^{LE}(q^2)&=&{3  m_\tau M_B \pi \sqrt{q^2} \over 2(2 m^2_\tau +q^2)}
 { {\rm Re}[ c^*_{10}(c_9 M_B + 2 c_7 m_b)] \over \left[|c_ {10}|^2 M_B^2 +
|c_9 M_B + 2 c_7 m_b|^2\right]}  \,\,\, .  \label{atkscet}
\eea
This is a  remarkable observation, which renders  the polarization asymmetries  important quantities
to measure even in this  exclusive mode.
Notice that in the derivation of the Large Energy relations one neglects terms of
 ${\cal O}(M^2_L / M^2_B)$ where $M_L$ is the light meson mass
 (in our case $M_K$ and $M_{K^*}$ considered below).
Consistently, we neglect such terms  and the strange quark mass in the derivation of 
${\cal A}_L^{LE}$ and ${\cal A}_T^{LE}$. 

\begin{figure}[ht]
\begin{center}
\includegraphics[width=0.45\textwidth] {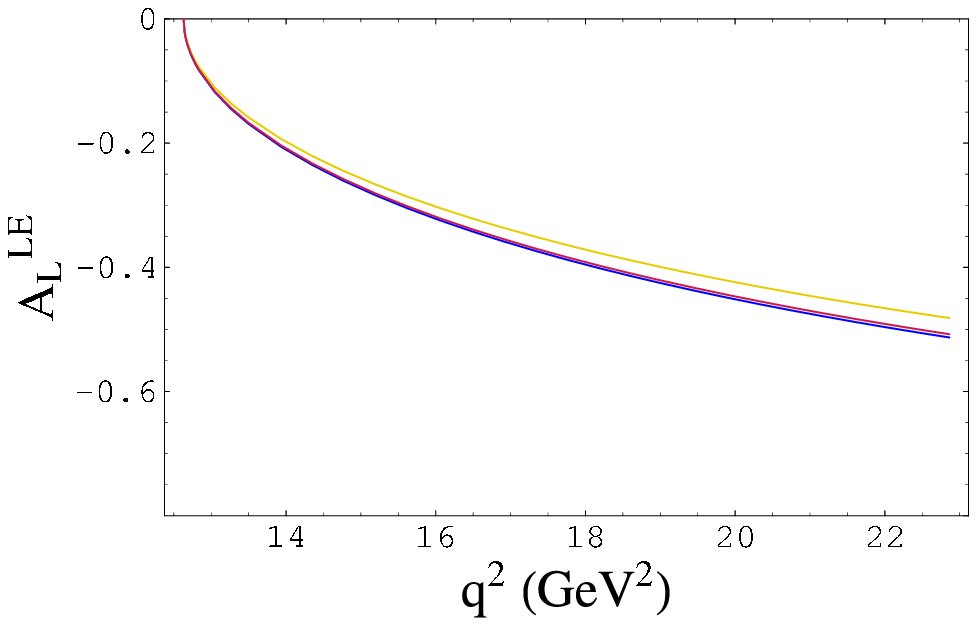} \hspace{0.5cm}
 \includegraphics[width=0.45\textwidth] {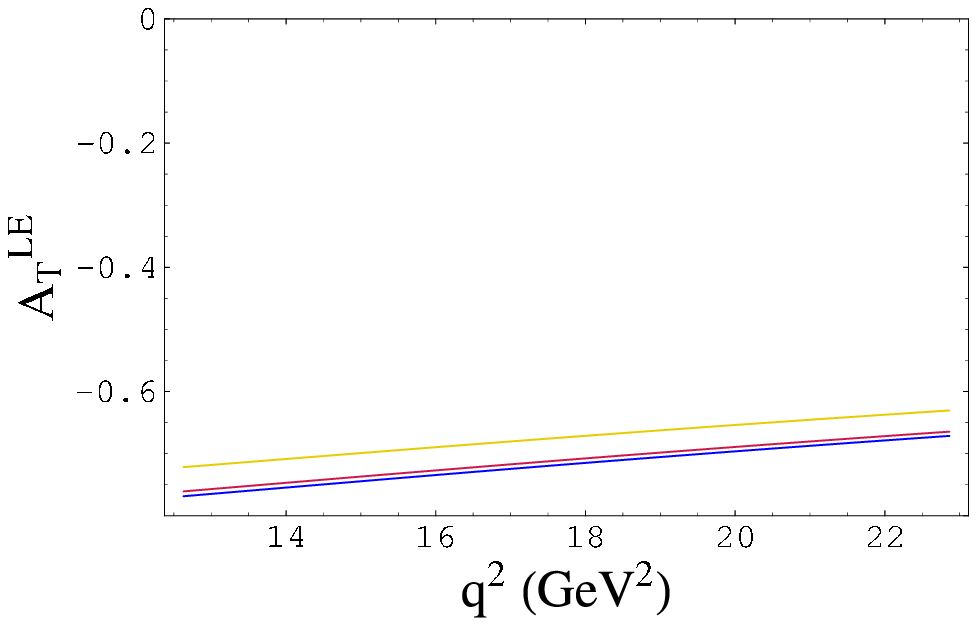}
\end{center}
\caption{\baselineskip=15pt Longitudinal   (left) and transverse (right) $\tau^-$ polarization
asymmetry in $B \to K \tau^+ \tau^-$  obtained using   the large energy limit
 relations for the hadronic matrix elements.
 The    blue curves refer to   SM,
 the intermediate (red) ones to  $1/R=500$ GeV, the light (yellow) ones to $1/R=200$ GeV.}
  \vspace*{1.0cm}
\label{fig:scetBK}
\end{figure}

The results obtained using eqs.(\ref{alkscet}),  (\ref{atkscet})
are depicted in fig.\ref{fig:scetBK}, where we have extrapolated to the full $q^2$ range
 the formulae obtained in the Large Energy limit
 which are strictly valid for small values of the momentum transfer.
 The systematic decrease (in absolute value) of the polarization asymmetries when $\dd 1/R$
 decreases is evident, and it is confirmed that the transverse polarization asymmetry is more sensitive
 to the universal extra dimension effects.
%

\section{Mode $B \to K^* \tau^+ \tau^-$}\label{sec:BKstar}
\begin{figure}[b]
\begin{center}
\includegraphics[width=0.45\textwidth] {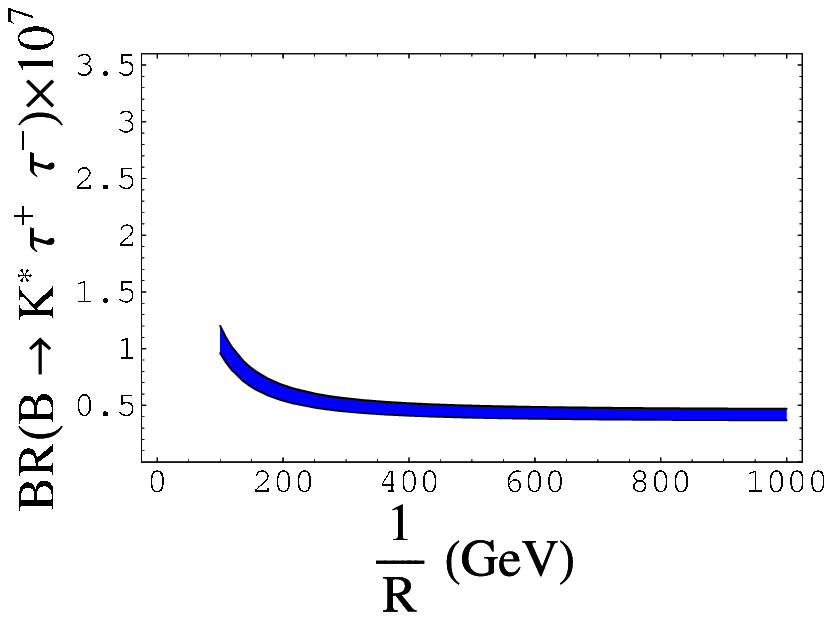} \hspace{0.5cm}
 \includegraphics[width=0.45\textwidth] {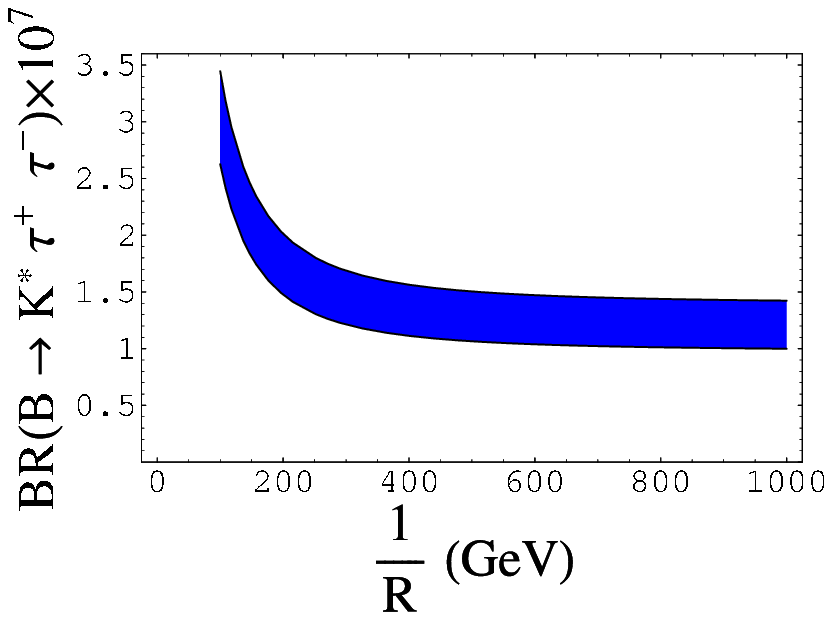}
\end{center}
\caption{\baselineskip=15pt Branching fractions of $B \to K^*
\tau^+ \tau^-$
 obtained using  set A   (left) and B  (right) of form factors, versus the compactification parameter
 $\dd 1/R$.} \vspace*{1.0cm}
\label{brkstartau}
\end{figure}

For $B \to K^* \tau^+ \tau^-$ the parameterization of the hadronic matrix elements involves more
 form factors:
\begin{eqnarray}
<K^*(p^\prime,\epsilon)|{\bar s} \gamma_\mu (1-\gamma_5) b
|B(p)>&=& \epsilon_{\mu \nu \alpha \beta} \epsilon^{* \nu}
p^\alpha p^{\prime \beta}
{ 2 V(q^2) \over M_B + M_{K^*}}  \nonumber \\
&-& i \left [ \epsilon^*_\mu (M_B + M_{K^*}) A_1(q^2) -
(\epsilon^* \cdot q) (p+p')_\mu  {A_2(q^2) \over (M_B + M_{K^*}) }
\right. \nonumber \\
&-& \left. (\epsilon^* \cdot q) {2 M_{K^*} \over q^2}
\big(A_3(q^2) - A_0(q^2)\big) q_\mu \right ] \label{a1}
\end{eqnarray}
\begin{eqnarray}
<K^*(p^\prime,\epsilon)|{\bar s} \sigma_{\mu \nu} q^\nu
{(1+\gamma_5) \over 2} b |B(p)>&=& i \epsilon_{\mu \nu \alpha
\beta} \epsilon^{* \nu} p^\alpha p^{\prime \beta}
\; 2 \; T_1(q^2)  + \nonumber \\
&+&  \Big[ \epsilon^*_\mu (M_B^2 - M^2_{K^*})  -
(\epsilon^* \cdot q) (p+p')_\mu \Big] \; T_2(q^2) \nonumber \\
&+& (\epsilon^* \cdot q) \left [ q_\mu - {q^2 \over M_B^2 -
M^2_{K^*}} (p + p')_\mu \right ] \; T_3(q^2)  \; ; \nonumber \\
\label{t1}
\end{eqnarray}
\noindent $A_3$ can be written as a combination of $A_1$ and
$A_2$:
\begin{equation}
A_3(q^2) = {M_B + M_{K^*} \over 2 M_{K^*}}  A_1(q^2) - {M_B -
M_{K^*} \over 2 M_{K^*}}  A_2(q^2)
\end{equation}
with the condition $ A_3(0) = A_0(0)$; the identity $\displaystyle
\sigma_{\mu \nu} \gamma_5 = - {i \over 2}
 \epsilon_{\mu \nu \alpha \beta} \sigma^{\alpha \beta}$
($ \epsilon_{0 1 2 3}=+1$) implies that $T_1(0) = T_2(0)$. The expression for the
differential decay rate:
\be {d\Gamma(q^2) \over d q^2}=
{G_F^2|V_{tb}V_{ts}^*|^2 \alpha^2 \over 2^{11} \pi^5}
{\lambda^{1/2}(M_B^2,M_{K^*}^2,q^2) \over M_B^3}\sqrt{1-{4
m_\tau^2 \over q^2}}{1 \over 3 m_{K^*}^2 q^2}g(q^2) \ee
involves the function $g(s)$:
 \bea
g(s)&=&24 |D_0|^2 m_\tau^2 M_{K^*}^2\lambda+8M_{K^*}^2s\,
\lambda[(2
m_\tau^2+s)|A|^2-(4m_\tau^2-s)|C|^2] \nonumber \\
&+&\lambda \left[(2
m_\tau^2+s)\big|B_1+(M_B^2-M_{K^*}^2-s)B_2\big|^2
-(4m_\tau^2-s)\big|D_1+(M_B^2-M_{K^*}^2-s)D_2\big|^2 \right] \nonumber \\
&+&4 M_{K^*}^2 s\, [(2 m_\tau^2+s)(3|B_1|^2- \lambda
|B_2|^2)-(4m_\tau^2-s)(3|D_1|^2- \lambda |D_2|^2) ]
\label{denakstar} \eea ($\lambda=\lambda(M_B^2,M_{K^*}^2,q^2)$),
where the terms $A$,$C$,$B_1$,$B_2$,$D_1$, $D_2$, $D_0$ contain
short distance coefficients as well as  form factors: \bea
A&=&{c_7 \over q^2}\,4\, (m_b+m_s)\, T_1(q^2)+c_9\,{V(q^2) \over M_B +M_{K^*}} \nn \\
C&=& c_{10}\, {V(q^2) \over M_B+M_{K^*}} \nn \\
B_1&=&{c_7 \over q^2}\, 4\, (m_b-m_s)\, T_2(q^2)(M_B^2-M_{K^*}^2)+ c_9\,A_1(q^2) ( M_B + M_{K^*}) \nn \\
B_2&=&- \left [{c_7 \over q^2}\, 4\, (m_b-m_s)\, \left (
T_2(q^2)+q^2
{T_3(q^2) \over (M_B^2-M_{K^*}^2)} \right )+ c_9 {A_2(q^2)\over M_B + M_{K^*}} \right ] \label{b2}\\
D_1&= &c_{10}\,A_1(q^2)\,(M_B+M_{K^*}) \nn \\
D_2&=&- c_{10} \,{A_2(q^2) \over M_B+M_{K^*}} \nn \\
D_0&=& c_{10} \,A_0(q^2)  \; . \nn
\eea
The $B \to K^* \tau^+ \tau^- $
branching fraction predicted within the Standard Model,  for  the two set of form factors,  is:
\be
BR(B \to K^* \tau^+ \tau^-)= \Bigg \{
\begin{array}{c}(4.1\pm 0.5) \,\,\, 10^{-8}  \,\,\, {\rm (set \,\, A)} \\ (1.2 \pm 0.2)\,\,\, 10^{-7}  \,\,\, {\rm (set \,\, B)\,\,\, ,} \end{array}
\ee
therefore  a remarkable hadronic uncertainty affects the decay rates,  analogously to the  case of
corresponding decays into light leptons ($e,\mu$) where at present the experimental results
are not able to distinguish between the two models  \cite{noi}.  Concerning the
dependence of the branching fractions on the compactification parameter $\dd 1/R$, it
is depicted in fig.\ref{brkstartau}: analogously to the case of $B \to K \tau^+ \tau^-$, a fraction
larger than $2 \times 10^{-7}$ is not compatible with the Standard Model result, so that
this eventual observation would put the bound $\dd {1 \over R} \leq 250$ GeV.

\begin{figure}[ht]
\begin{center}
\includegraphics[width=0.45\textwidth] {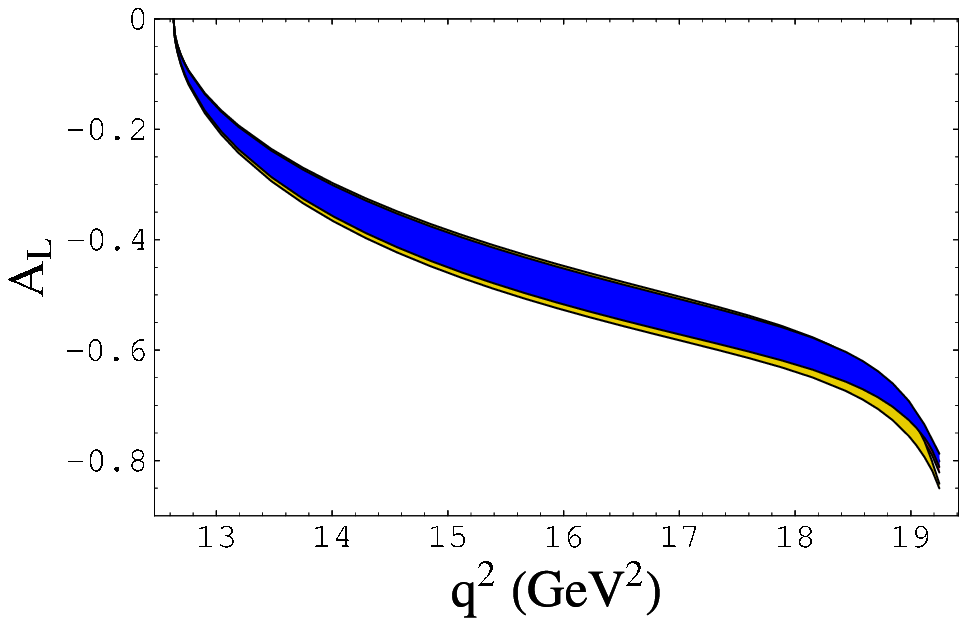} \hspace{0.5cm}
 \includegraphics[width=0.45\textwidth] {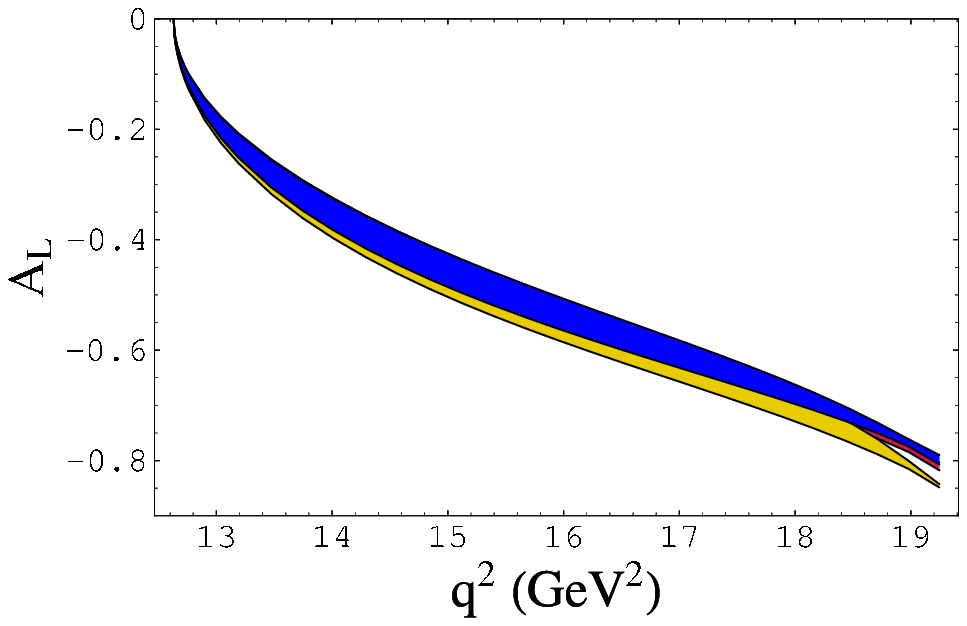}\\\vspace{0.5cm}
 \includegraphics[width=0.45\textwidth] {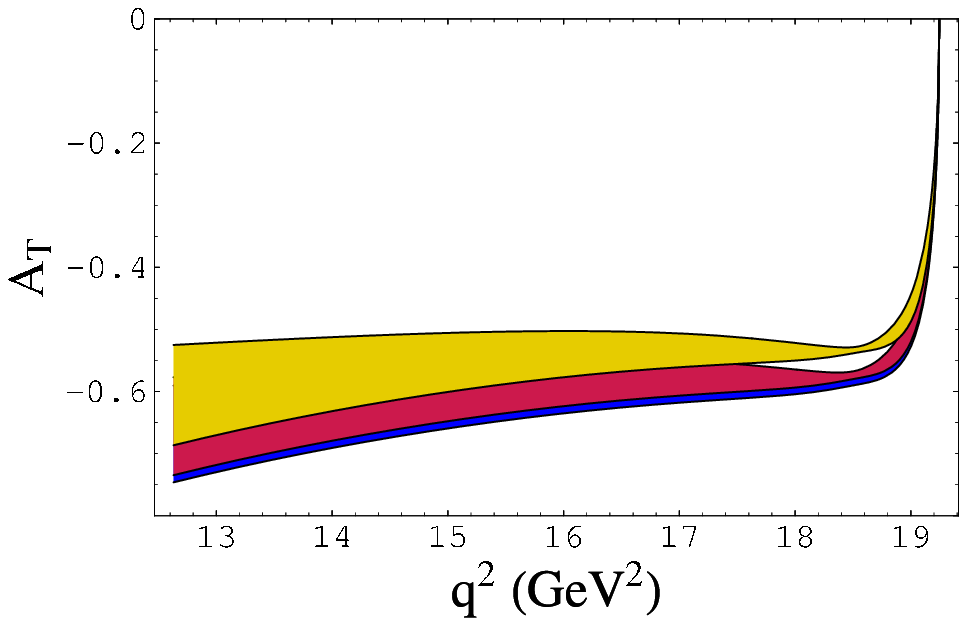} \hspace{0.5cm}
 \includegraphics[width=0.45\textwidth] {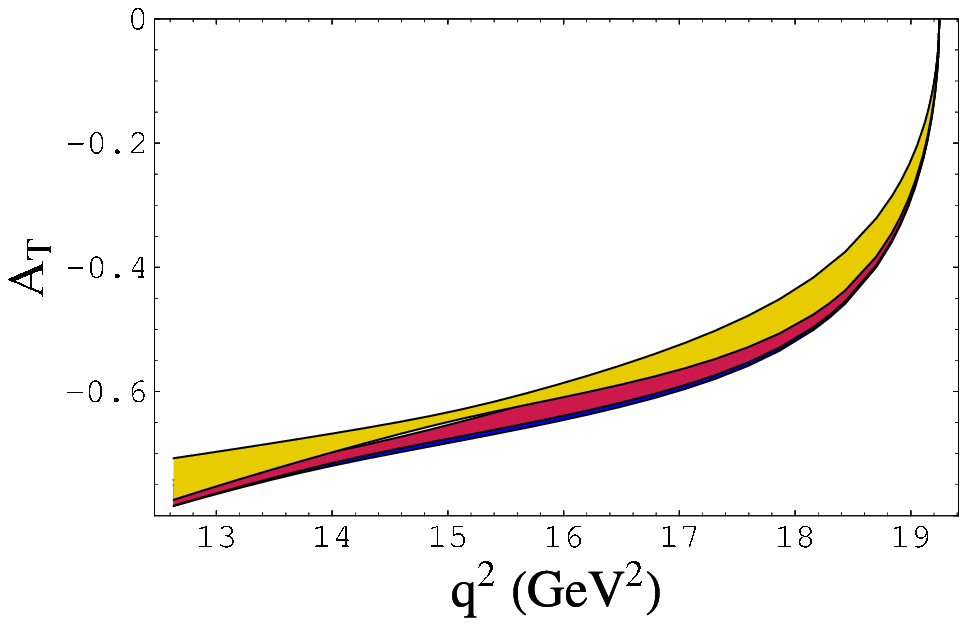}
\end{center}
\caption{\baselineskip=15pt Longitudinal (up) and transverse (down)  $\tau^-$ polarization
asymmetry in $B \to K^* \tau^+ \tau^-$
 obtained using  set A   (left) and B  (right) of form factors.
 The  dark (blue) region is obtained in  SM;
 the intermediate (red) one for $1/R=500$ GeV, the light (yellow) one for $1/R=200$ GeV.} \vspace*{1.0cm}
\label{asymLkstar}
\end{figure}

Concerning the  $\tau^-$ longitudinal ${\cal A}_L$ and transverse  ${\cal A}_T$  polarization
asymmetries, their expressions:
\bea {\cal A}_L(q^2)&=& 2 q^2
\sqrt{1-{4 m_\tau^2 \over q^2}}{1 \over g(q^2)} 
 \Bigg\{ 8 M_{K^*}^2 q^2{\rm Re}[B_1D_1^*+\lambda AC^*] \nn \\
&+&{\rm Re}\bigg[ [(M_B^2-M_{K^*}^2-q^2)B_1+\lambda
B_2] [(M_B^2-M_{K^*}^2-q^2)D_1^*+\lambda D_2^* ] \bigg]
\Bigg\}\nn \\  \label{along-star}
 \eea
\be {\cal A}_T(q^2)=3 \pi m_\tau M_{K^*}{\sqrt{q^2} \lambda^{1/2}
\over g(q^2)} \left[-4{\rm Re}[A \, B_1^*]\, M_{K^*}q^2+{\rm
Re}\big[D_0 \, [B_1^*(M_B^2-M_{K^*}^2-q^2)+B_2^* \lambda]\big]
\right]\label{atrans-star} \ee involve the combinations of Wilson
coefficients and form factors in (\ref{b2}) and the function in
(\ref{denakstar}). The results for the two models, depicted in
fig.\ref{asymLkstar} and computed in the Standard Model and for
two values of the compactification radius, show that the effect of
the universal extra dimension is  modest. The longitudinal
polarization asymmetry ${\cal A}_L$, which vanishes at $q^2=4
m_\tau^2$, has its largest (absolute) value for  the largest
momentum transfer and is practically insensitive to
$\dd 1/R$. Sensitivity is higher for the  transverse polarization
asymmetry ${\cal A}_T$, which decreases (in absolute value) by
nearly $15 \%$ with the decrease  of   $\dd 1/R$ down to
$\dd {1 \over R}=200$ GeV; such an effect is maximal in the low
$q^2$ range.
%

 It is interesting to consider  the large energy limit of the light vector meson
 in the final state, since also in this case relations can be obtained among the form factors
 \cite{scet}. In particular, only  two functions $\xi_\perp$ and
$\xi_\parallel$ appear as independent nonperturbative quantities in terms of which the
$B \to K^*$ form factors can be written:
\bea
V(q^2)&=& -i {M_B +M_{K^*} \over M_B}\xi_\perp(E)\nonumber \\
A_1(q^2)&=& -i{2E \over M_B +M_{K^*}} \xi_\perp (E) \nonumber\\
A_2(q^2)&=& i {M_B \over M_B -M_{K^*}}(\xi_\parallel(E)-\xi_\perp(E) ) \nonumber\\
A_0(q^2)&=&-i {E \over M_{K^*}} \xi_\parallel(E)  \label{scetV}\\
T_1(q^2)&=&-{i \over 2} \xi_\perp (E) \nonumber \\
T_2(q^2) &=& -i{E \over M_B}\xi_\perp (E) \nonumber\\
T_3(q^2)&=&{i \over 2}(\xi_\parallel(E)-\xi_\perp (E))\nn
\eea
with $E={M_B\over 2}(1+{M_K^{*2} \over M_B^2}-{q^2\over M_B^2})$.
Using these relations,  the expressions for the longitudinal and transverse $\tau^-$ polarization asymmetries turn out to be equal to the corresponding ones in the case of $B \to K \tau^+ \tau^-$,
i.e. eqs.(\ref{alkscet}), (\ref{atkscet}).
Therefore,  in the large energy limit the dependence on the hadronic functions $\xi_\perp$ and
$\xi_\parallel$ cancels out  in ${\cal A}_L$ and ${\cal A}_T$. Such
a noticeable feature allows us to consider  also these two asymmetries as  important
quantities for testing the Standard Model and possible extra dimension effects.

\section{$K^*$ helicity fractions in $B \to K^* \ell^+ \ell^-$}\label{sec:BKstarhelicity}
This last Section of our study is devoted
to the discussion of other interesting observables in $B \to K^* \ell^+ \ell^-$ transitions,  i.e. the helicity fractions of the vector meson $K^*$ produced in the final state. The interest in these
observables  has been prompted by the recent
 BaBar Collaboration  measurement of  the longitudinal $K^*$
helicity fraction $f_L$ in the modes  $B \to K^* e^+ e^-, \, K^* \mu^+ \mu^-$  \cite{Aubert:2006vb}.
The measurement has been  done in   two bins of  momentum transfer $q^2$, with the result:
\bea
f_L&=&0.77^{+0.63}_{-0.30}\pm 0.07 \hskip 1cm 0.1
\leq q^2 \leq 8.41 \,\,\, GeV^2 \,\,\,\,  \nonumber\\
f_L&=&0.51^{+0.22}_{-0.25}\pm 0.08 \hskip 1cm  q^2 \geq 10.24  \,\,\, GeV^2
\label{flexp} \,\,\, , \eea
while the average value of $f_L$ in the full $q^2$ range is:
\be
f_L=0.63^{+0.18}_{-0.19}\pm 0.05 \hskip 1cm
q^2 \geq 0.1 \,\,\, GeV^2 \,\,\,\, .
\ee

The expressions  of  $B \to K^* \ell^+ \ell^-$ differential decay widths with  $K^*$ longitudinally  (L)  or
transversely  ($\pm$) polarized read as follows:

\bea {d\Gamma_L(q^2) \over d q^2}&=& {G_F^2|V_{tb}V_{ts}^*|^2
\alpha^2 \over 2^{11} \pi^5} {\lambda^{1/2}(M_B^2,M_{K^*}^2,q^2)
\over M_B^3}\sqrt{1-{4 m_\ell^2 \over q^2}}{1 \over 3}\, A_L
\nonumber \\
{d\Gamma_+(q^2) \over d q^2}&=& {G_F^2|V_{tb}V_{ts}^*|^2 \alpha^2
\over 2^{11} \pi^5} {\lambda^{1/2}(M_B^2,M_{K^*}^2,q^2) \over
M_B^3}\sqrt{1-{4 m_\ell^2 \over q^2}}{4 \over 3}\, A_+
 \\
{d\Gamma_-(q^2) \over d q^2}&=& {G_F^2|V_{tb}V_{ts}^*|^2 \alpha^2
\over 2^{11} \pi^5} {\lambda^{1/2}(M_B^2,M_{K^*}^2,q^2) \over
M_B^3}\sqrt{1-{4 m_\ell^2 \over q^2}}{4 \over 3}\, A_- \nn \eea
with
\bea A_{L}&=&{1 \over q^2 \, M_{K^*}^2} \bigg\{24  |D_0|^2 m_\ell^2
M_{K^*}^2\lambda
+ (2 m_\ell^2+q^2)\left|B_1(M_B^2-M_{K^*}^2-q^2) + B_2
\lambda\right|^2\nonumber
\\ &+&(q^2-4 m_\ell^2) \left|D_1(M_B^2-M_{K^*}^2-q^2) +
D_2 \lambda\right|^2\bigg\} \label{along} \eea \noindent and \bea
A_-&=&(q^2-4m_\ell^2) |D_1+\lambda^{1/2}C|^2+(q^2+2m_\ell^2) |B_1+\lambda^{1/2}A|^2
\nonumber
\\
A_+&=&(q^2-4m_\ell^2) |D_1-\lambda^{1/2}C|^2+(q^2+2m_\ell^2) |B_1-\lambda^{1/2}A|^2
\label{transverse}\eea
\noindent where $\lambda=\lambda(M_B^2,M_{K^*}^2,q^2)$ and the combinations in (\ref{b2})
of Wilson coefficients and form factors have been used. The various helicity fractions are
defined as:
\bea
f_L (q^2) &=&{ {d\Gamma_L(q^2) /d q^2} \over {d\Gamma(q^2) /d q^2}}  \nonumber \\
f_\pm (q^2) &=&{ {d\Gamma_\pm(q^2) /d q^2} \over {d\Gamma(q^2) /d q^2}} \\
f_T(q^2)&=&f_+(q^2)+f_-(q^2) \nn
 \label{fvarie} \eea
 so that $f_L(q^2)+f_+(q^2)+f_-(q^2)=f_L(q^2)+f_T(q^2)=1$ for each value of $q^2$.
 
\begin{figure}[t]
\begin{center}
\includegraphics[width=0.45\textwidth] {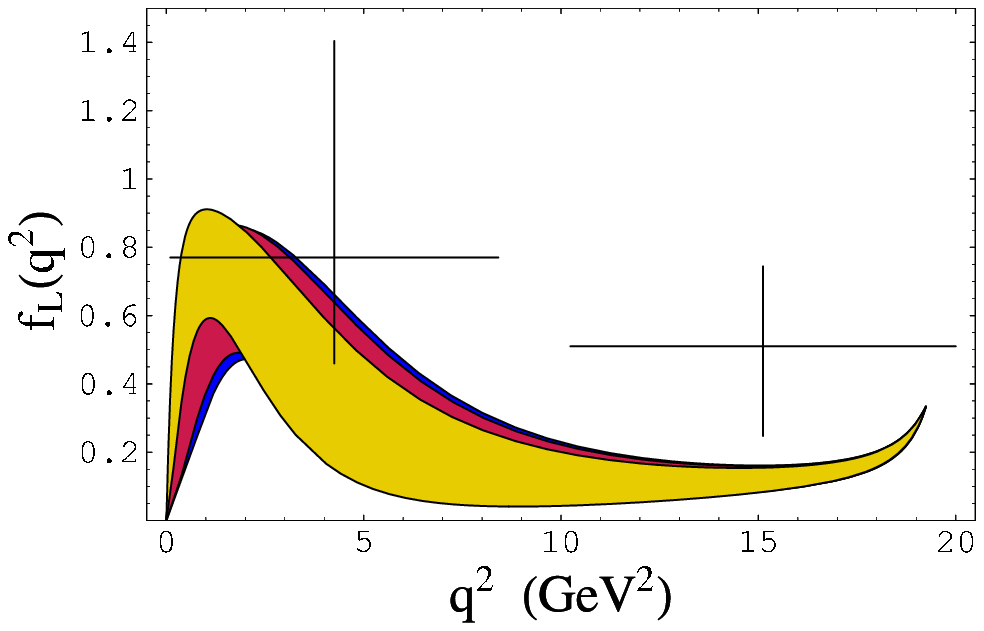} \hspace{0.5cm}
 \includegraphics[width=0.45\textwidth] {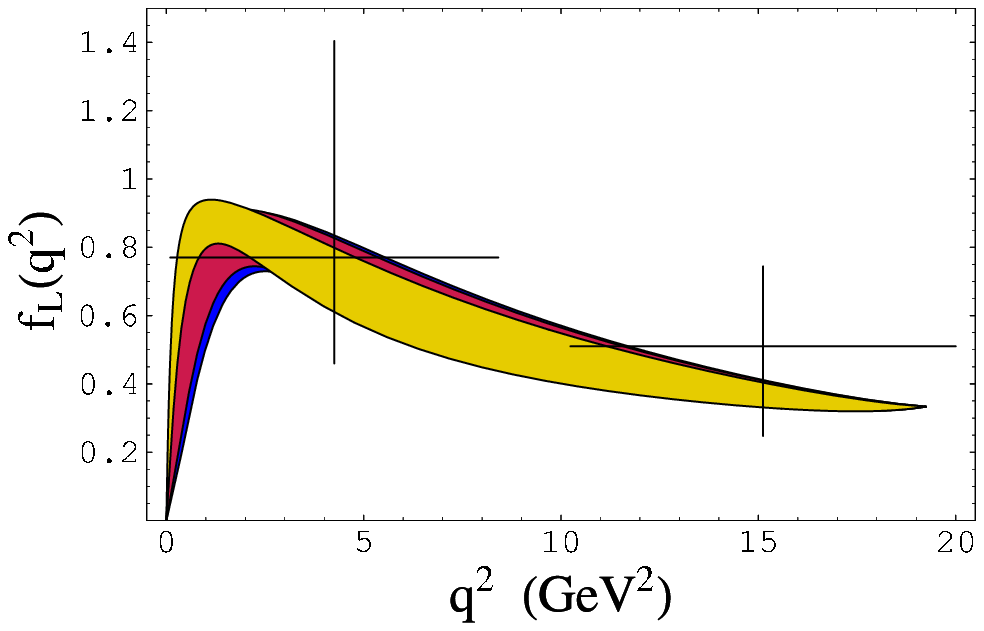}\\ \vspace{0.5cm}
 \includegraphics[width=0.45\textwidth] {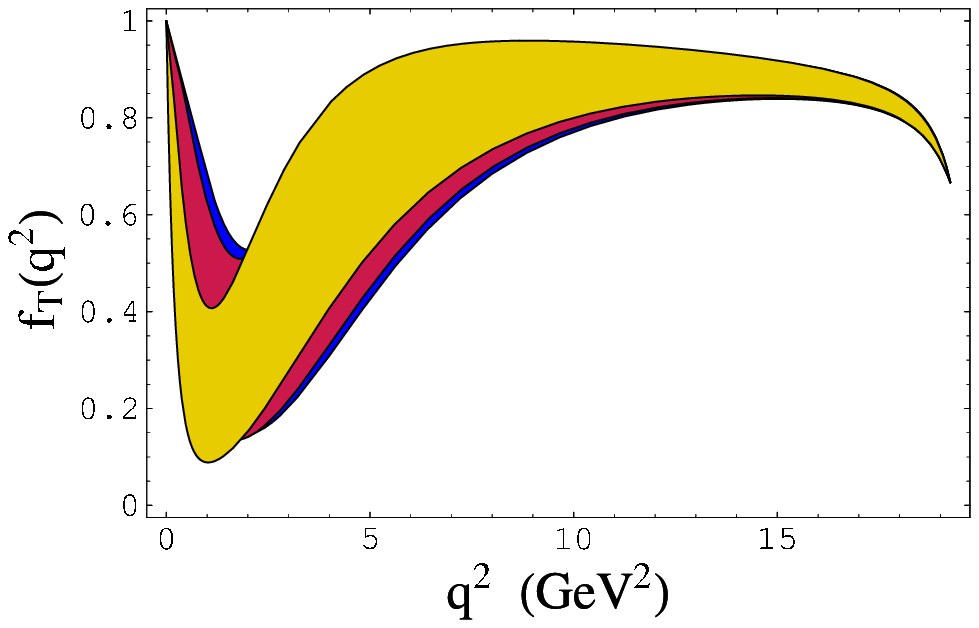} \hspace{0.5cm}
 \includegraphics[width=0.45\textwidth] {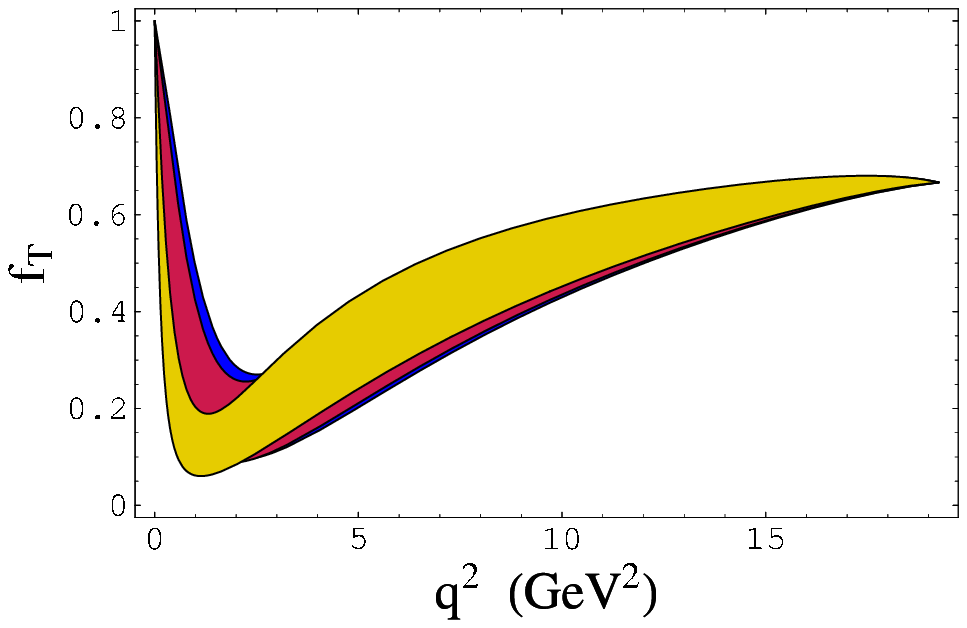}
\end{center}
\caption{\baselineskip=15pt  Longitudinal (up) and transverse (down) $K^*$
helicity fraction in $B \to K^* \ell^+ \ell^-$ ($\ell=\mu,e$)
 obtained using  set A   (left) and B  (right) of form factors.
 The  dark (blue) region is obtained in  SM;
 the intermediate (red) one for $1/R=500$ GeV, the light (yellow) one for $1/R=200$ GeV.
 The  two measured values of longitudinal polarization in two bins of $q^2$:
$0.1\leq q^2 \leq 8.41 \,\,\, {\rm GeV}^2$ and
$ q^2 \geq 10.24  \,\,\, {\rm GeV}^2$
 are shown in the top panels.} \vspace*{1.0cm}
\label{fLdif}
\end{figure}

 These helicity  fractions have been considered in  SM
and in some extensions e.g. in \cite{Aliev:2001fc};  the
 results obtained in the Standard Model and for two values of $\dd 1/R$  are shown in fig.\ref{fLdif}
 in the case $m_\ell=0$.
For this observable, the  dependence on the compactification parameter $\dd1/R$ is mild for  $\dd {1 \over R} \geq 200$ GeV compared to the hadronic uncertainty. The results  for the longitudinal $K^*$ fraction $f_L(q^2)$,
for the two sets of form factors,
agree with the experimental data within the experimental uncertainties: however, in the
high $q^2$ range there is better  agreement if  set $B$ is used, as one can
appreciate looking at fig.\ref{fLdif}.  A measurement of the
tranverse $K^*$ helicity fraction could contribute to discriminate between the two sets, mainly considering the high range of momentum transfer where the results from the two sets are rather different.

\begin{figure}[ht]
\begin{center}
\includegraphics[width=0.45\textwidth] {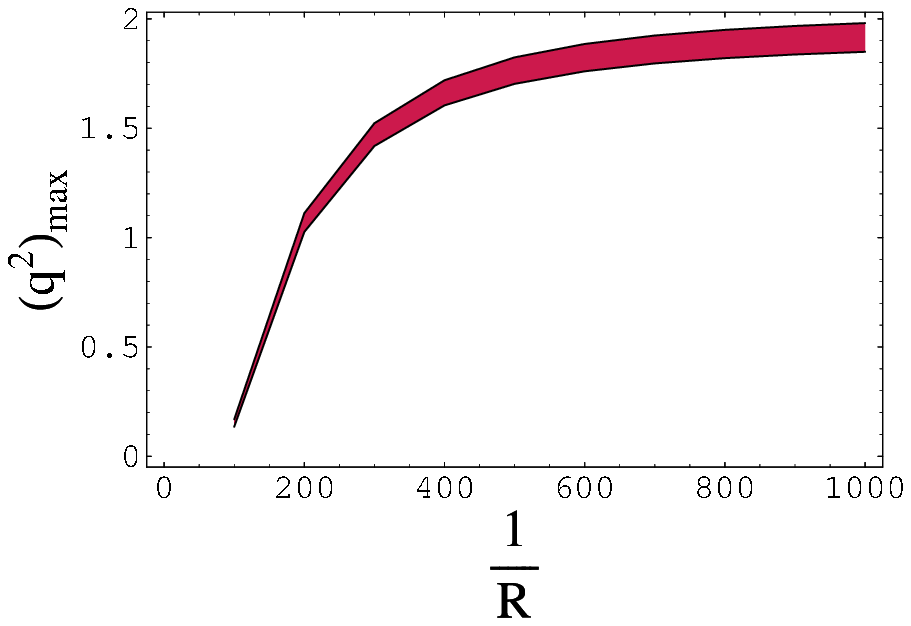} \hspace{0.5cm}
 \includegraphics[width=0.45\textwidth] {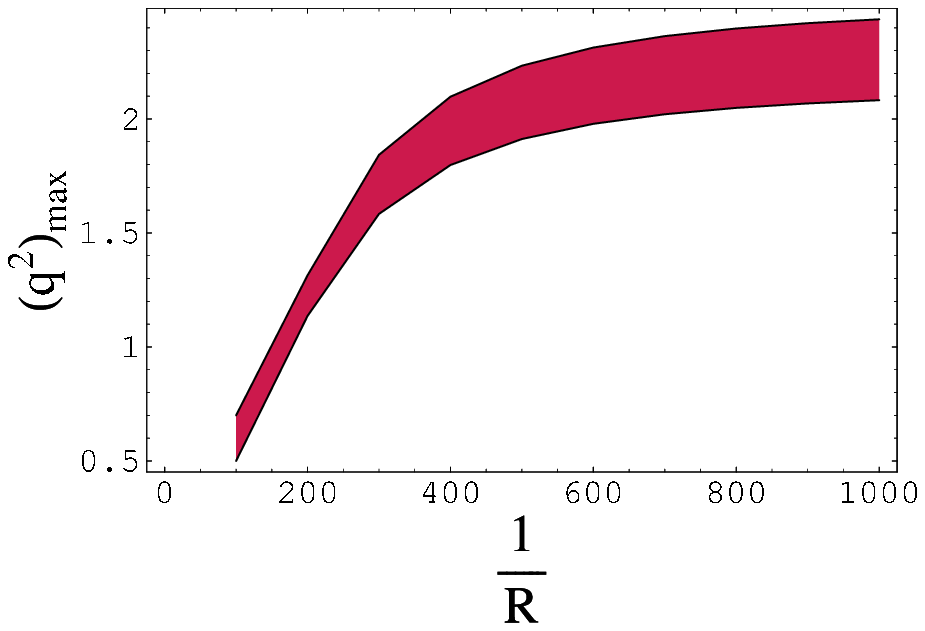}
\end{center}
\caption{\baselineskip=15pt  Value of momentum transfer $q^2_{max}$ (in GeV$^2$) corresponding to
 the maximum of  $K^*$ longitudinal
 helicity amplitude $f_L(q^2)$ in $B \to K^* \ell^+ \ell^-$
($\ell=\mu,e$)
 obtained using  set A   (left) and B  (right) of form factors, versus the compactification parameter
 $\dd 1/R$. } \vspace*{1.0cm}
\label{maxfL}
\end{figure}

Although the general dependence on the compactification parameter is modest, there is a quantity
in the longitudinal $K^*$ fraction $f_L(q^2)$ where sensitivity to $\dd 1/R$ is higher:
the value of momentum transfer $q^2=q^2_{max}$ where the longitudinal fraction has a maximum.
The dependence of this point on $\dd 1/R$ is depicted in
fig.\ref{maxfL}, where  it is shown  how the position of the maximum
of $f_L$ varies with $\dd 1/R$:  the position of the maximum is shifted, when
$\dd 1/R$ decreases, towards smaller values, therefore a precise measurement of the position $q^2_{max}$
can be used to constrain  $\dd 1/R$. For example, $q^2_{max}\simeq 1$ 
GeV$^2$ is not compatible with the Standard Model, but with  an ED scenario with  $\dd {1 \over R}\simeq 200$ GeV.

\begin{figure}[ht]
\begin{center}
\includegraphics[width=0.45\textwidth] {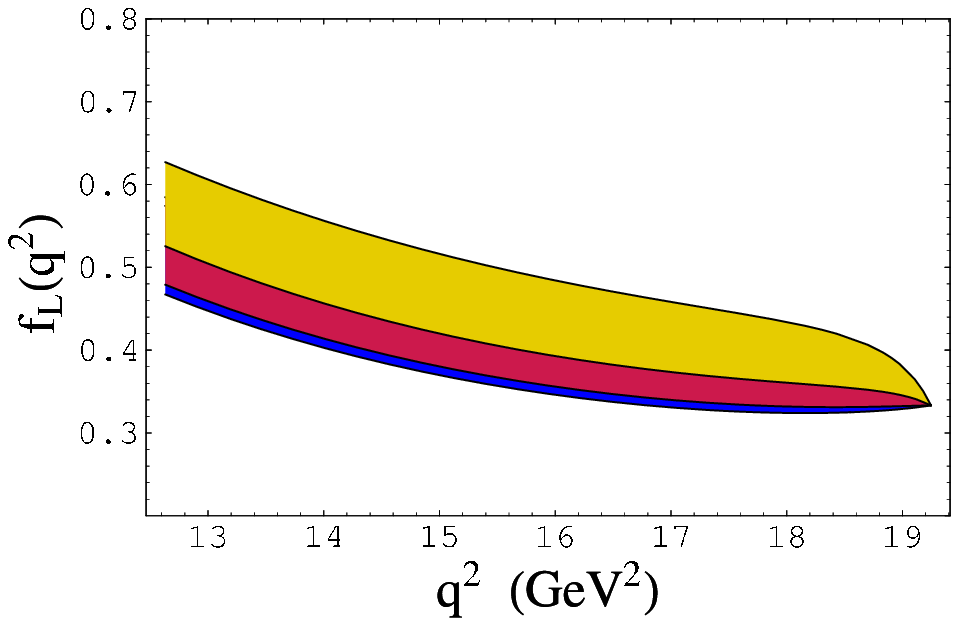} \hspace{0.5cm}
 \includegraphics[width=0.45\textwidth] {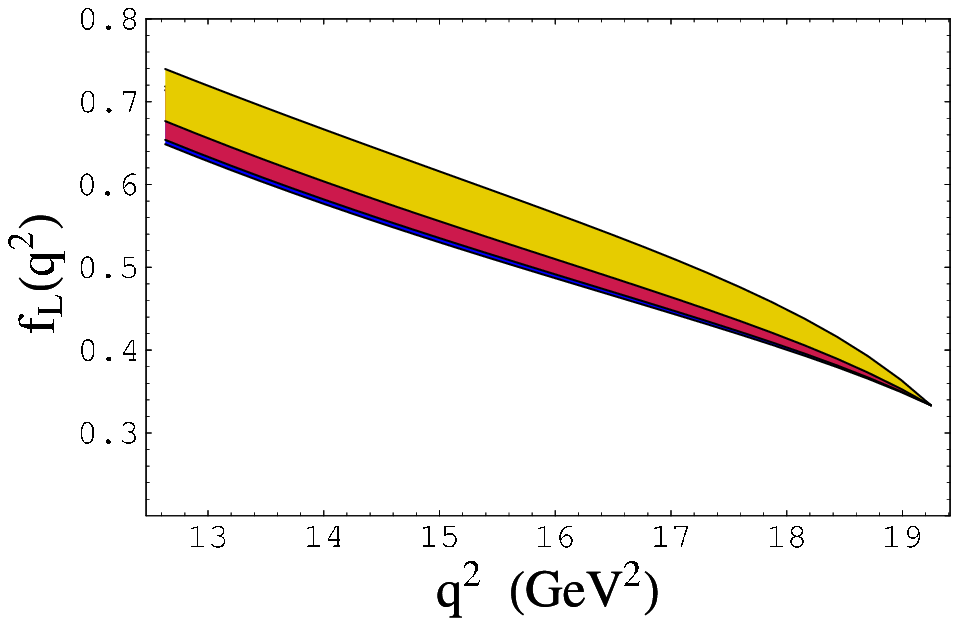}\\ \vspace{0.5cm}
 \includegraphics[width=0.45\textwidth] {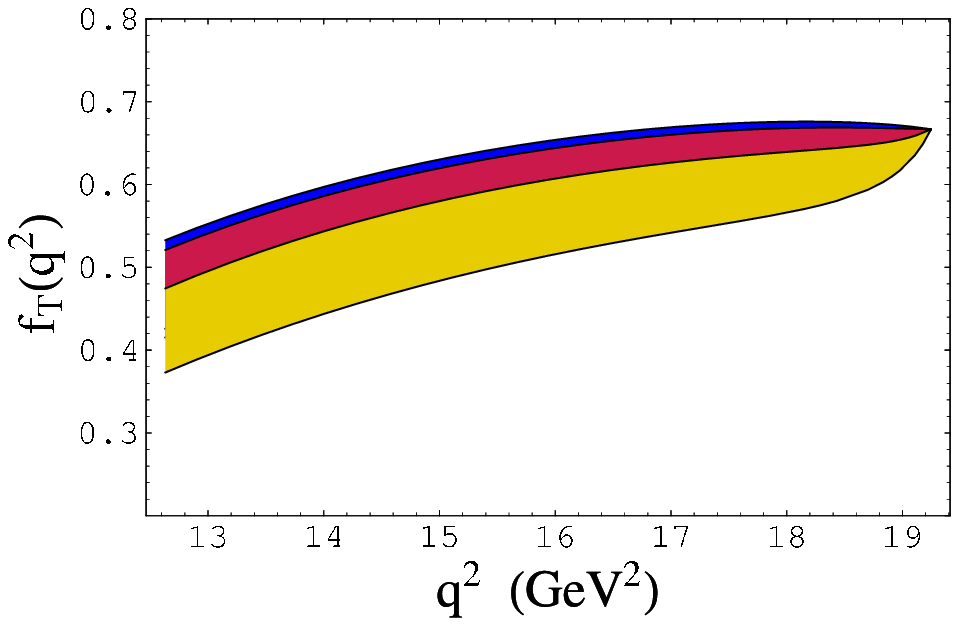} \hspace{0.5cm}
 \includegraphics[width=0.45\textwidth] {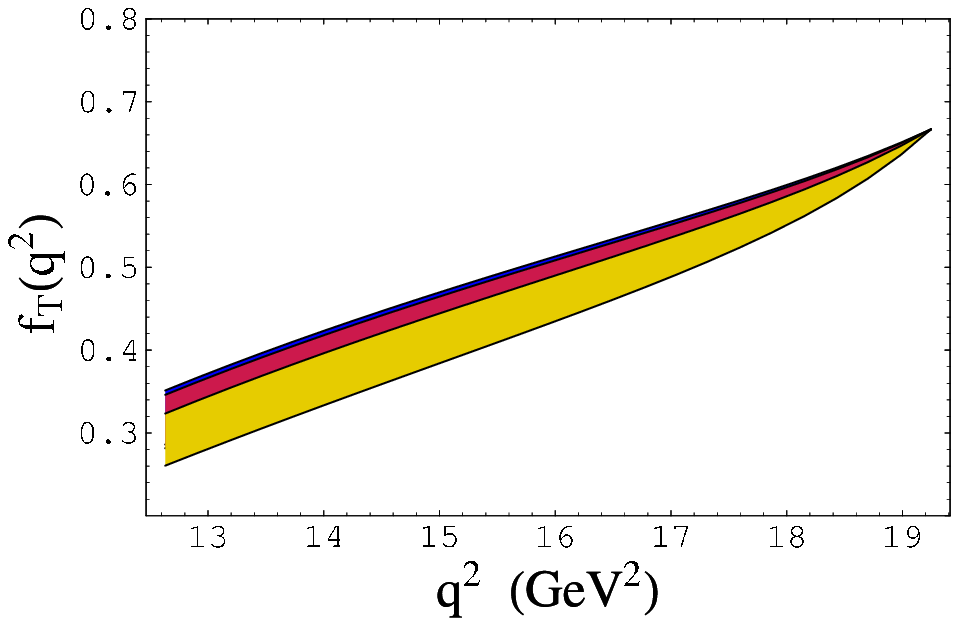}
\end{center}
\caption{\baselineskip=15pt Longitudinal  (up) and transverse (down) $K^*$ helicity fraction
in $B \to K^* \tau^+ \tau^-$
 obtained using  set A   (left) and B  (right) of form factors. The meaning of the shadowing (colors)
 is the same as in fig.\ref{fLdif}.
 } \vspace*{1.0cm}
\label{fLtau}
\end{figure}

For  $B \to K^* \tau^+ \tau^-$ transitions,
the predicted $K^*$ longitudinal and transverse helicity fractions
are shown in fig.\ref{fLtau}. In this case the dependence of the fractions 
on $q^2$ is monotonic, and the maximum (minimum) value of  $f_L$ ($f_T$) is obtained
for the smallest momentum transfered.
%

It is possible to express the $K^*$ helicity amplitudes in the Large Energy limit,  using
the relations in (\ref{scetV}):
\bea A_L^{LE} &=& {1 \over q^2 M_{K^*}^2}{(M_B^2-q^2)^4 \over
M_B^2} (2 m_\ell^2+q^2)
(|c_{9,\parallel}|^2+|c_{10}|^2)\xi^2_\parallel(q^2) \nonumber
 \\
A_-^{LE}&=&4{(M_B^2-q^2)^2 \over M_B^2}\left[ (2 m_\ell^2+q^2)
|c_{9,\perp}|^2+(-4 m_\ell^2+q^2)|c_{10}|^2 \right] \xi^2_\perp
\nonumber \\
A_+^{LE}&=&0 \eea
 \noindent where we have  introduced the following combinations of
 Wilson coefficients:
 \bea
 c_{9,\parallel}&=& c_9+{2 m_b c_7 \over M_B} \nonumber \\
c_{9,\perp}&=& c_9+ {2  c_7 m_b M_B\over q^2}  \,\,\, . \eea
As for  lepton polarization asymmetries,  we neglet terms of
${\cal O}(M^2_{K^*}/M^2_B)$ apart from the phase space. Actually,  $A_+$ 
turns out to be proportional to $M^2_{K^*}$. The dependence on the hadronic matrix element now
is reduced to the explicit dependence on the two form factors $\xi_\parallel$ and $\xi_\perp$ for which 
 some determination must be provided. 

We conclude this Section observing that the measurement of  the $K^*$  helicity distributions is  possible 
at  $B$ factories  and  represents  an important completion  of the study
of rare $B \to K^* \ell^+ \ell^-$ transitions.

\section{Conclusions}\label{sec:conclusions}

In this study we have investigated how spin observables in $B \to X_s \tau^+ \tau^-$ and
$B \to K^{(*)} \tau^+ \tau^-$ transitions can be
used to provide a bound to the radius $R$ of the compactified extra dimension in the ACD model, 
considering also the results expected in the Standard Model to which the ACD model 
reduces in the  $R \to 0$ limit. We have found that the branching fractions are sensitive to 
$\dd 1/R$, so that can be used to provide us with such an information. We have also found that
the dependence  of the $\tau^-$ polarization asymmetries on  $\dd 1/R$
is mild but still observable, the most sensitive ones
being the transverse asymmetries. During our investigation we have shown that 
in the exclusive modes the polarization asymmetries are
free of hadronic uncertainties if one considers the Large Energy limit for the light hadron
in the final state,  an important observation as far as the use
of these observables for testing the Standard Model is concerned. Finally, we have considered
the $K^*$ helicity fractions, for which some results are already available
when the leptons in the final state are $\ell=e,\mu$.

Present and future experimental  activities   aim at
testing   the Standard Model and  the possible extensions  with their  numerous  aspects.
Among the  New Physics scenarios, models with extra dimensions  at first glance would seem  to belong to the class of the exotic ones: on the contrary,  we have shown that these models can be severely  constrained  using experimental data already available or attainable in the near future; such
tests  represent a further step  in the search of the ultimate theory of fundamental interactions.

\vspace*{1cm} \noindent
{\bf Acnowledgments}
 One of us (PC) thanks CPhT,  \'Ecole
Polytechnique,  Palaiseau, for kind hospitality during the completion of this work.

 
\end{document}